\documentclass[aps,prx,twocolumn,notitlepage,amsmath,amstex,amssymb,superscriptaddress,longbibliography]{revtex4-1}

\usepackage{lmodern} 
\usepackage{microtype} 
\usepackage[english]{babel}

\usepackage{color,graphicx}

\usepackage{dsfont,url}
\usepackage{slashed,cancel}
\usepackage[usenames,dvipsnames]{xcolor}
\usepackage[colorlinks=true, linkcolor=black, citecolor=ForestGreen]{hyperref}
\usepackage{mdframed}

\usepackage{physics,tikz}

\usepackage[caption=false]{subfig}

\usetikzlibrary{decorations.markings,calc}

\newcommand{\RM}[1]{{\color{Orchid}{[{\bf RM}: #1]}}}

 \renewcommand{\RM}[1]{{}}

\newcommand{\new}[1]{\textcolor{black}{#1}}

\begin{document}

\title{\bf 
Theory of Out-of-Time-Ordered Transport
}

\author{Ruchira Mishra}
\affiliation{Leinweber Institute for Theoretical Physics \& James Franck Institute, University of Chicago, Chicago, IL 60637, USA}
\author{Jiaozi Wang}
\affiliation{University of Osnabr\"uck, Department of Mathematics/Computer Science/Physics, D-49076 Osnabr\"uck, Germany}
\author{Silvia Pappalardi}
\affiliation{Institut f\"ur Theoretische Physik, Universit\"at zu K\"oln, Z\"ulpicher Straße 77, 50937 K\"oln, Germany}
\author{Luca V. Delacr\'etaz}
\affiliation{Leinweber Institute for Theoretical Physics \& James Franck Institute, University of Chicago, Chicago, IL 60637, USA}

\begin{abstract}
We construct an effective field theory (EFT) that captures the universal behavior of out-of-time-order correlators (OTOCs) at late times in generic quantum many-body systems with conservation laws. The EFT hinges on a generalization of the strong-to-weak spontaneous symmetry breaking pattern adapted to out-of-time-order observables, and reduces to conventional fluctuating hydrodynamics when time-ordered observables are probed. We use the EFT to explain different power-law behavior observed in OTOCs at late times, and show that many OTOCs are entirely fixed by conventional transport data \new{at leading order}. Nevertheless, we show that a specific combination of OTOCs is sensitive to novel transport parameters, not visible in regular time-ordered correlators. We test our predictions in Hamiltonian and Floquet spin chains in one spatial dimension.

\end{abstract}

\maketitle

\section{Introduction}

The thermalization dynamics of quantum many-body systems has been a subject of active research in the past decades \cite{Deutsch:1991nq,Srednicki:1994ce,Rigol:2008zz,DAlessio:2016rwt}. Despite the complexity of the microscopic dynamics, universal behavior typically emerges at late times, including in the form of hydrodynamics, random matrix behavior, and patterns of entanglement or operator growth. Fluctuating hydrodynamics, in particular, offers one of the few controlled tools to make quantitative predictions in chaotic quantum many-body systems with conservation laws. The success of an effectively classical hydrodynamic description obscures to some extent the microscopic quantum complexity and extensive entanglement that allowed it to emerge. Understanding whether the hydrodynamic regime hides signatures of the underlying quantum dynamics is an important outlying question.

Recently, hydrodynamic behavior has been observed in quantum observables that do not have an obvious classical counterpart. These include the spectral form factor \cite{Friedman:2019gyi,Winer:2020gdp}, entanglement growth \cite{Rakovszky:2019oht,Huang:2019hts}, and out-of-time-order correlators (OTOCs) \cite{Rakovszky:2017qit,Khemani:2017nda}. While it is not surprising that hydrodynamic signatures pervade observables beyond conventional linear response and transport, understanding and predicting these signatures quantitatively requires a quantum description of fluctuating hydrodynamics going beyond the conventional Martin-Siggia-Rose (MSR) formalism \cite{Martin:1973zz,dominicis1976techniques,Janssen:1976qag,PhysRevA.16.732}.

The framework that will allow us to go beyond classical observables is the recent formulation of fluctuating hydrodynamics as an effective field theory (EFT) on a Schwinger-Keldysh contour \cite{Grozdanov:2013dba,Crossley:2015evo,Haehl:2015foa,Jensen:2018hse,Liu:2018kfw}. While most applications of these EFTs to date led to results that could have been obtained from the classic MSR formalism, their construction is naturally extendable to capture purely quantum observables. In this paper, we focus on one of the simplest such observables, out-of-time-order correlators (OTOCs), with the goal of uncovering their universal behavior in generic thermalizing systems.

Much of the early work on OTOCs was in the context of semiclassical, large $N$, or weakly coupled systems, where microscopic calculations and effective descriptions showed that OTOCs feature exponential Lyapunov growth inside a butterfly cone \cite{larkin1969quasiclassical,Shenker:2013pqa,Maldacena:2015waa,Aleiner:2016eni,Blake:2017ris,Grozdanov:2018atb,Blake:2018leo,Keselman:2020fmo,lerose2020bridging, Choi:2023mab,Gao:2023wun, garcia2022out,PhysRevB.108.L241106}. Although Lyapunov growth does not survive away from these simplifying limits, numerical studies have shown that butterfly cones may be more generic \cite{Kukuljan:2017xag,vonKeyserlingk:2017dyr,Khemani:2017nda,Rakovszky:2017qit, Nahum2018operator, xu2022scrambling}. The operator growth dynamics that is responsible for the butterfly front happens for local quantum many-body systems whether or not conserved quantities are present. Refs.~\cite{Khemani:2017nda,Rakovszky:2017qit} studied systems with conservation laws, and found an interesting interplay of this `information dynamics' with conserved densities.

The goal of this paper is to establish a controlled theoretical framework that captures OTOCs in generic quantum many-body systems with conservation laws. Are hydrodynamic signatures in OTOCs quantitatively related to those in more conventional correlators? Can one {\em predict} the asymptotic, late time OTOCs from knowledge of time-ordered correlators (TOCs) only? We will find that this is the case to a certain extent---regular OTOCs are entirely fixed, asymptotically, in terms of time-ordered correlators---however, certain linear combinations of OTOCs reveal novel transport parameters that are invisible in time-ordered correlators alone.

We start by describing a simple heuristic argument in Sec.~\ref{sec_heuristic} that explains the power laws observed in previous numerical studies. In  Sec.~\ref{sec_EFT_construction}, we construct the EFT that provides a controlled approach to OTOCs at late times. This construction reveals the existence of novel transport parameters even in simple diffusive systems, that are only visible in OTOCs.  Sec.~\ref{sec_EFT_properties} further explores these transport parameters and discusses the properties of the EFT. Finally, we compare our predictions to Hamiltonian and Floquet systems with conservation laws in Sec.~\ref{sec_numerics}.

\section{Heuristic scaling argument}\label{sec_heuristic}

\begin{figure}[t]
	\includegraphics[width=1\columnwidth]{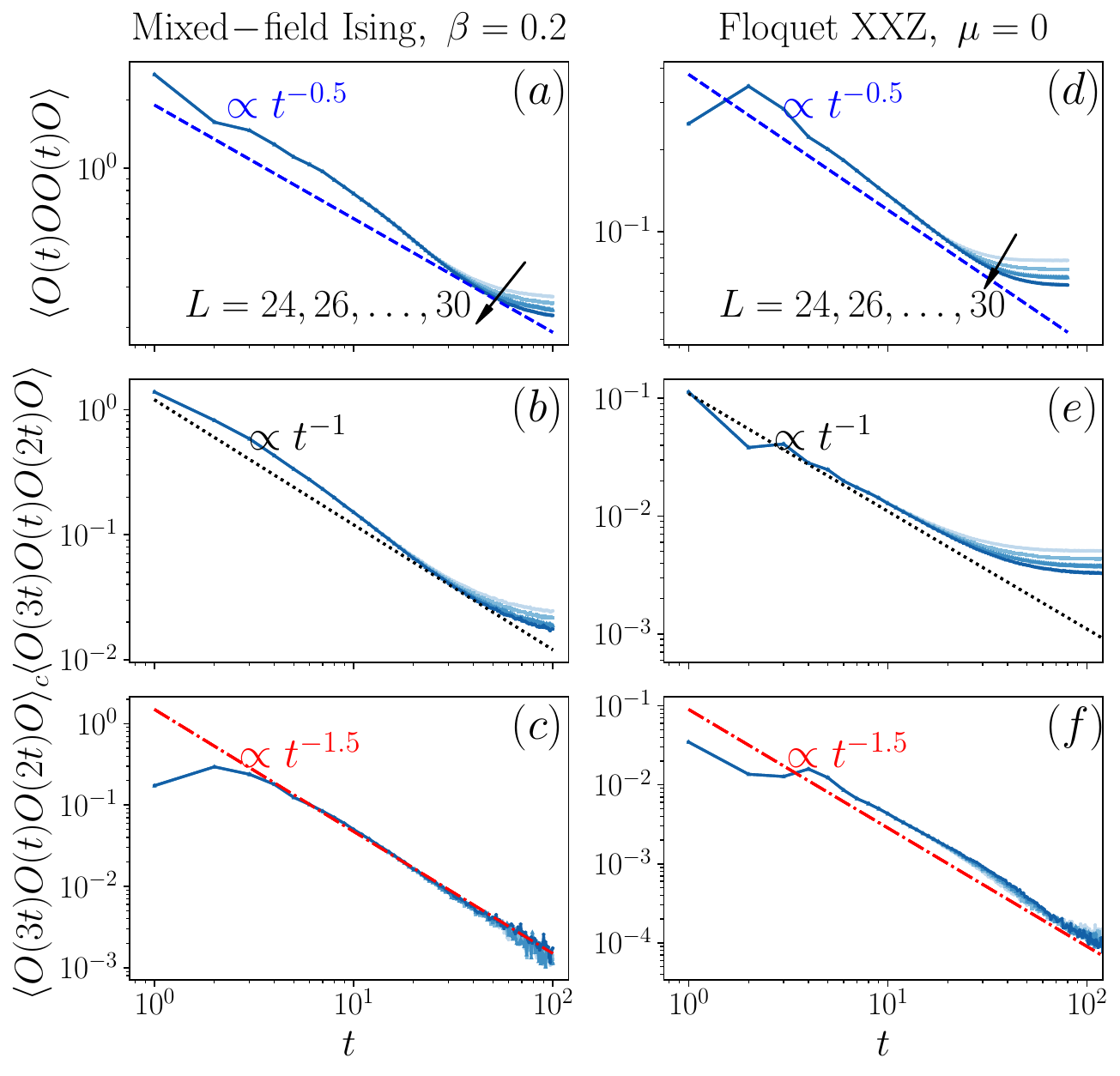}
\caption{\small Out-of-time-ordered correlators in the mixed-field Ising model at $\beta = 0.2$ [(a)(b)(c)]  and in the Floquet XXZ model [(d)(e)(f)] at $\mu=0.0$. In both models, the operator $\mathcal{O} = n - \langle n \rangle$, where $n$ is the conserved local density. \new{The OTOC at coincident times $\langle O(t)OO(t)O\rangle$, and (dis)connected OTOC at separate times $\langle O(3t)O(t)O(2t)O\rangle$ and $\langle O(3t)O(t)O(2t)O\rangle_c$ obey three different power laws as predicted by the EFT.} 
}
\label{fig_powerlaws}
\end{figure}

OTOCs in systems with conservation laws were observed to have diffusive tails in \cite{Rakovszky:2017qit,Khemani:2017nda}. Fig.~\ref{fig_powerlaws} reproduces and generalizes their results to OTOCs at non-coincident points, revealing several different power laws: $1/t^{1/2}$, $1/t$, and $1/t^{3/2}$. Before presenting the detailed construction of the EFT that will quantitatively predict OTOCs at late times, we describe a simple heuristic argument that captures this observed scaling behavior. At late times, in a hydrodynamic regime, operators that overlap with conserved densities approximately commute: $[A,B]\sim \tau \omega AB$, where $\omega$ is the frequency scale of the probe, and $\tau$ is an intrinsic time scale of the hydrodynamic description ($\tau\omega \ll 1$ in the hydrodynamic regime). This statement will be sharpened later---in fact we will see it is a prediction of the EFT---but for now we simply note that it is consistent with the fluctuation-dissipation relation for two-point functions $\Im G^R(\omega) = \tanh \frac{\beta\omega}{2} G^S(\omega) \simeq \frac{\beta\omega}{2} G^S(\omega)$, where in this case the intrinsic time scale is set by inverse temperature $\tau=\beta \equiv  1/ T$ (we set $\hbar=1$).

Thus, OTOCs are approximately equal to TOCs. For example,
\begin{equation}\label{eq_OTOC_ex}
\langle \mathcal{O}(t_3)\mathcal{O}(t_1)\mathcal{O}(t_2)\mathcal{O}(0)\rangle_\beta
	\simeq \langle \mathcal{O}(t_3)\mathcal{O}(t_2)\mathcal{O}(t_1)\mathcal{O}(0)\rangle_\beta\, , 
\end{equation}
where $\langle \cdot \rangle_\beta \equiv \Tr (\rho_\beta\, \cdot \, )$ denotes expectation values in the thermal state $\rho_\beta = e^{-\beta H} / \Tr (e^{-\beta H})$.
The corrections to the equality have a relative $1/t$ suppression at large time separations $t = \min_{ij}|t_i - t_j|$ (this error estimate is conservative and can be improved when the operators are also spatially separated).

To leading order, OTOCs of the form \eqref{eq_OTOC_ex} can thus be obtained from regular time-ordered higher-point functions. Nonlinear TOCs in the hydrodynamic regime were studied in \cite{Delacretaz:2023ypv}, and are characterized by a simple scaling behavior which we briefly review. Consider a hydrodynamic regime described by the scaling $\omega\sim q^z$, with $z=2$ corresponding to diffusion. The charge density two-point function behaves as
\begin{equation}
\langle n(t) n(0)\rangle_\beta = \frac{1}{t^{d/z}} + \cdots\, ,
\end{equation}
so that charge fluctuations scale as $\delta n\sim 1/t^{d/(2z)}\sim q^{d/2}$. Nonlinear response or higher-point functions arise from non-Gaussianities in the EFT, which are suppressed by additional powers of fluctuations $\delta n$. Since a connected $N$-point function requires $N-2$ such insertions, its overall scaling is
\begin{equation}\label{eq_Nptscaling}
\langle \mathcal{O}(t_N) \cdots \mathcal{O}(t_1)\rangle_{c,\beta}
	\sim \frac{1}{t^{d(N-1)/z}}\, .
\end{equation}
Here, $\mathcal O$ is any operator overlapping with the diffusing density $n$. Other operators may also feature power-law decay due to overlap with composite hydrodynamic fields ($\partial_x n,\, (\delta n)^2$, etc.) \cite{Glorioso:2020loc,Matthies:2024lqx}; for example, in the mixed-field Ising model {\em every} operator of finite support is expected to feature polynomial decay \cite{Delacretaz:2023pxm}.

Setting $d=1$, $z=2$ (diffusion), and $N=4$, Eq.~\eqref{eq_Nptscaling} explains the $1/t^{3/2}$ scaling observed in Fig.~\ref{fig_powerlaws}. Implicit in this statement is the fact that regular (non-connected) correlators approximately factorize, because the hydrodynamic EFT is weakly coupled (even if the underlying microscopics is strongly coupled!). Thus, for $N$ even, a non-connected $N$-point function is approximately equal to a product of $N/2$ two-point functions, which scales as $\sim (1/t^{d/z})^{N/2}$. Setting again $d=1$, $z=2$ and $N=4$ reproduces the $1/t$ scaling observed in Fig.~\ref{fig_powerlaws}.

We must still explain the $1/\sqrt{t}$ scaling of the `coincident' OTOC, observed in Fig.~\ref{fig_powerlaws} as well as \cite{Rakovszky:2017qit}. The results above break down if two operators are taken closer to each other than the UV cutoff of the hydrodynamic EFT. Beyond this limit, pairs of nearby operators can be viewed as a single composite operator, and expanded in an operator product expansion of the EFT \cite{Delacretaz:2023ypv}. The 4-point OTOC therefore reduces to a single {\em two}-point function, whose scaling indeed agrees with the observed behavior \footnote{Note that when the microscopic description is a field theory, we do not expect its OPE to apply to the OTOC (see, e.g., \cite{Caron-Huot:2022lff}). However, because the OTOC is a regular correlator of our Schwinger-Keldysh EFT (Sec.~\ref{sec_EFT_construction}), the OPE of the EFT applies.}.

These heuristic results---and in particular Eq.~\eqref{eq_OTOC_ex}, which shows that OTOCs approximately reduce to TOCs at late times---already point to where novel universal phenomena may be found: in the {\em difference} between OTOCs and TOCs. Correlators involving commutators can capture the difference between OTOCs and TOCs, for example
\begin{equation}\label{eq_otoc_commutator}
G_{\rm otoc}(t_i) \equiv
	\langle \mathcal{O}(t_3)[\mathcal{O}(t_2),\mathcal{O}(t_1)]\mathcal{O}(0)\rangle_\beta\, .
\end{equation}
As will be discussed below, this observable reveals novel transport parameters that are invisible in time-ordered correlators or conventional response functions. \new{These transport parameters also give subleading contributions to other OTOCs like \eqref{eq_OTOC_ex} at late times, but the fact that they control the leading order behavior of \eqref{eq_otoc_commutator} makes them most physically relevant and straightforward to extract in numerics.}

\section{EFT for out-of-time-ordered transport}\label{sec_EFT_construction}

\subsection{Generating functional for OTOCs}

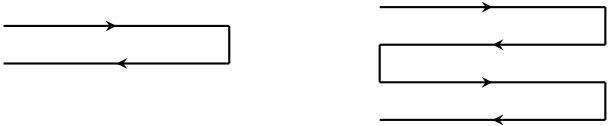
\begin{figure}[t]
    \centering
    \begin{tikzpicture}[
        scale=0.5,
        >=stealth,
        decoration={markings, mark=at position 0.5 with {\arrow{stealth}}}
    ]

        \draw[thick, postaction={decorate}] (-3,2.5) -- (3,2.5);
        \draw[thick] (3,2.5) -- (3,1.5);
        \draw[thick, postaction={decorate}] (3,1.5) -- (-3,1.5);
     
      \begin{scope}[xshift=10cm]
      \draw[thick, postaction={decorate}] (-3,3) -- (3,3);
      \draw[thick] (3,3) -- (3,2);
      \draw[thick, postaction={decorate}] (3,2) -- (-3,2);
      \draw[thick] (-3,2) -- (-3,1);
      \draw[thick, postaction={decorate}] (-3,1) -- (3,1);
      \draw[thick] (3,1) -- (3,0);
      \draw[thick, postaction={decorate}] (3,0) -- (-3,0);
 \end{scope}

    \end{tikzpicture}

    \caption{The 1-CTP (`closed-time path') and 2-CTP contours.}
    \label{fig_contours}
\end{figure}

Consider a quantum many-body system with time-evolution generated by a Hamiltonian $H_0$. Our construction will also apply to Floquet dynamics and local unitary circuits with conservation laws, but we use the notation of Hamiltonian evolution for concreteness. We assume the system has at least one continuous global symmetry, which could be time translation invariance, leading to the conservation of a current
\begin{equation}
\partial_\mu j^\mu = 0\, .
\end{equation}
It will be useful to couple this current to external sources: 
\begin{equation}
H_0 \to H[A](t) = H_0(t) - \int d^d x A_\mu(t,\vec x) j^\mu(t,\vec x)\, .
\end{equation}
One could also similarly source other operators. We will drop the $\mu$ index and let $A$ collectively denote any source that is considered.
The time-evolution unitary is then
\begin{equation}
U[A](t_f,t_i) = \mathcal{T} e^{-i \int_{t_i}^{t_f} dt \, H[A](t)}\, , 
\end{equation}
where $\mathcal{T}$ denotes time-ordering. The conventional Schwinger-Keldysh generating functional is given by
\begin{equation}\label{eq_Z2}
Z_2[A_1,A_2]
	\equiv \Tr \left( \rho  U_2^\dagger U_1\right)\, ,
\end{equation}
where we use the short-hand notation
\begin{equation}
U_i \equiv U[A_i](\infty,-\infty)\, .
\end{equation}
This functional can be used to generate any time-ordered correlator of currents and densities. We will instead consider a generalization with four `switchbacks':
\begin{equation}\label{eq_Z4}
Z_4[A_1,A_2,A_3,A_4]
	\equiv \Tr \left( \rho U_4^\dagger U_3 U_2^\dagger U_1\right)\, .
\end{equation}
The corresponding closed-time paths (CTP) are illustrated in Fig.~\ref{fig_contours}. $Z_4$ can generate any 4-point OTOC, as well as a large class of higher-point functions. Our goal is to find an EFT that controls the asymptotic late time expansion of $Z_4$.

This generating functional can also be expressed in the path integral formalism in terms of the microscopic action $S[\psi]$, where $\psi$ denotes the microscopic fields or lattice degrees of freedom:
\begin{equation}\label{eq_Z4_PI}
Z_4 = \int_{\rm BC} D\psi_{1,2,3,4}\,  e^{i (S[\psi_1]- S[\psi_2] + S[\psi_3] - S[\psi_4])}\, .
\end{equation}
The boundary conditions (BC) corresponding to the object \eqref{eq_Z4} are: $\psi_1(t_f) = \psi_2(t_f)$, $\psi_2(t_i) = \psi_3(t_i)$, and $\psi_3(t_f) = \psi_4(t_f)$,  and finally the path integral is weighted by the matrix element $\langle \psi_1(t_i) |\rho| \psi_4(t_i)\rangle$.

\subsection{Properties of the generating functional}\label{ssec_Z4_properties}

The generating functional \eqref{eq_Z4} satisfies a number of properties, which would be automatic in any microscopic realization. However, because the EFT constructed in this paper is not microscopic, these properties will have to be carefully imposed by hand to constrain the EFT.

\begin{enumerate}
\item
	{\em Collapse:} if sources on any two successive legs are identified (say, $A_1=A_2=A$), these two legs `collapse' $U_2^\dagger U_1 = 1$ and the result is independent of $A$. In equations:
	\begin{equation}\label{eq_collapse}
	\begin{split}
	&Z_4[A_1,A_2,A,A]=
	Z_4[A_1,A,A,A_2]\\
	&=
	Z_4[A,A,A_1,A_2]=
	Z_2[A_1,A_2]\, .
	\end{split}
	\end{equation}
\item
	{\em Unitarity} of the time evolution operator implies
	\begin{equation}\label{eq_unitarity}
	Z_4[A_1,A_2,A_3,A_4]^* = Z_4[A_4,A_3,A_2,A_1]\, .
	\end{equation}
\item
	{\em KMS:} the thermal state $\rho = e^{-\beta H} / \Tr e^{-\beta H}$ can be commuted across any number of unitaries, giving
	\begin{subequations}\label{eq_KMS}
	\begin{align}
	Z_4 &= Z_4[A^1(-t+i\beta),A^4(-t),A^3(-t),A^2(-t)]\, , \\
	Z_4 &= Z_4[A^3(t),A^4(t),A^1(t+i\beta),A^2(t+i\beta)]\,.
	\end{align}
	\end{subequations}
\end{enumerate}
In the second equation, $\rho$ was commuted across $U_1$ and $U_2^\dagger$. In the first equation, time-reversal symmetry was applied after commuting $\rho$ across $U_1$,  to recover a generating functional of the form \eqref{eq_Z4}. We note that time-reversal symmetry is not necessary to impose the KMS symmetries \cite{Wang:1998wg,Sieberer:2015hba,Haehl:2017eob}: one can instead define a reversed generating functional $\widetilde Z_4 = \Tr \left(\rho U_4 U_3^\dagger U_2 U_1^\dagger\right)$, and relate $Z_4$ to $\widetilde Z_4$ (in the EFT, the Wilsonian coefficients of both must be the same as they can both produce the same observables).

These constraints generalize similar properties of $Z_2$ that played an important role in recovering fluctuating hydrodynamics from an EFT on a Schwinger-Keldysh contour \cite{Crossley:2015evo, Haehl:2015foa, Jensen:2018hse}. Note that the collapse constraint \eqref{eq_collapse} in particular implies that the 2-CTP EFT reduces to the conventional 1-CTP one (fluctuating hydrodynamics). Thus, the 2-CTP EFT we will construct will also contain all the information of fluctuating hydrodynamics, as well as new transport parameters that can appear in OTOCs.

The generating functional also satisfies `partial collapse' constraints that are stronger than \eqref{eq_collapse}: if two consecutive sources are set to be equal past a certain time $t_o$ (say, $A_1 - A_2 \propto \theta(t_o - t)$), then the dependence of the generating functional on the source past $t_o$ drops out. This condition is sometimes referred to as a {\em latest time condition}, and is further discussed in App.~\ref{app_Z_to_S}. As we will see, this stronger condition will be automatically satisfied in the EFT.

\subsection{Symmetries of mixed state time evolution}

Let us take the microscopic system to be invariant under a global $U(1)$ symmetry for concreteness, $[H_0,Q] = 0$. There is a natural doubling of symmetries in the dynamics of mixed states: time evolution of a density matrix $\rho$ commutes with the action of a symmetry generator on either side $e^{i\alpha_1 Q} \rho$ or $\rho e^{-i\alpha_2 Q}$ of the density matrix \cite{Buca:2012zz}. This doubling was already used in \eqref{eq_Z2}, where the symmetry currents were coupled to independent gauge fields. 

In the generating function for OTOCs, there are further additional symmetries for each switchback. One can view the object being time evolved as an element $\rho\otimes \mathds 1 \in \mathcal (\mathcal H\otimes \mathcal H^*)^2$ of a quadrupled Hilbert space---a symmetry action on any of the 4 legs commutes with time evolution for a closed system. The dynamics is thus invariant under $U(1)^4$.

Now that we have discussed symmetries of the time evolution, we would like to know if  observables respect these symmetries. Consider an operator $\psi$ charged under the $U(1)$. Notice that charged thermal correlators generically do not vanish 
\begin{equation}
\Tr \rho \psi^\dagger \psi \neq 0\, , \quad
\Tr \rho \psi^\dagger \psi^\dagger \psi \psi \neq 0\,, \quad \cdots
\end{equation}
We have not specified the spacetime argument of the operators as it does not play an important role below; one can imagine that they are slightly separated in time.
In terms of correlation functions of the $\psi_i$ appearing in \eqref{eq_Z4_PI}, the non-vanishing of charged thermal correlators implies that, e.g.,
\begin{equation}
\langle \psi_{i_1}^\dagger\psi_{i_2}^\dagger \psi_{i_3}\psi_{i_4} \rangle
	\neq 0\, , \qquad \forall i_{n} \in \{1,2,3,4\}\, .
\end{equation}
We thus see that all of the symmetries $\psi_i \to e^{i\alpha_i}\psi_i$ are generically broken by observables, except for the diagonal $U(1)_{\rm diag}$ that acts on all legs in the same way, $\psi_i \to e^{i\alpha} \psi_i$. We will interpret this invariance of the observables only under the diagonal $U(1)_{\rm diag}$ as a spontaneous breaking of the symmetries (SSB). Note that the symmetries, and the SSB pattern, depend on the observable under consideration: for generating functionals $Z_2$, $Z_4$, or more generally $Z_{2n}$, the symmetry is $U(1)^{2n}$ and is expected to be spontaneously broken to the diagonal group:
\begin{equation}\label{eq_SSB}
U(1)^{2n} \xrightarrow{\ \rm SSB\ }
	U(1)_{\rm diag}\, .
\end{equation}
This statement generalizes to any symmetry group $G$, which can be non-abelian, higher-form, and include discrete factors; we generically expect thermal dynamics to realize the SSB pattern
\begin{equation}\label{eq_SSB_gen }
G^{2n} \xrightarrow{\ \rm SSB\ }
	G_{\rm diag}\, .
\end{equation}
For continuous symmetry groups, this spontaneous symmetry breaking (SSB) pattern will lead to $(2n-1)\dim G$ Nambu-Goldstone--like fields that nonlinearly realize the symmetry. This generalizes the proposal of viewing fluctuating hydrodynamics as a theory of Nambu-Goldstone (NG) bosons for the the strong-to-weak spontaneous symmetry breaking (SWSSB) of continuous symmetries in thermal states \cite{Ogunnaike:2023qyh,Akyuz:2023lsm}, which is the $n=1$ version of our statement (see \cite{Moudgalya:2023yon,Delacretaz:2023pxm,Gu:2024wgc,Huang:2024rml,Firat:2025upx} for further discussions on fluctuating hydrodynamics from the SWSSB perspective, and \cite{Lessa:2024gcw,Zhang:2024fpf,Zhou:2025bal} for studies of strong discrete or generalized symmetries and their breaking). The Nambu-Goldstone fields can be thought of as noise fields that accompany the density in fluctuating hydrodynamics, the winding history of OTOCs allowing for multiple noise realizations in our case.

\subsection{Degrees of freedom of the EFT}

The SSB pattern identified above immediately suggests the collective field content that governs the late time dynamics. First consider the symmetry breaking pattern $U(1)^2\to  U(1)_{\rm diag}$ that pertains to $Z_2$ and time-ordered correlators. If the entire $U(1)^2$ were spontaneously broken, the EFT would contain two NG fields
\begin{equation}
\phi_1,\,\phi_2\, .
\end{equation}
It is convenient to work with fields that are even and odd under interchange of the two legs (`Keldysh rotation') 
\begin{equation}\label{eq_KeldyshRotation_Z2}
\phi_r = \frac12 (\phi_1 + \phi_2)\, , \qquad
\phi_a = \phi_1 - \phi_2\, .
\end{equation}
Such an EFT describes the ordered (superfluid) phase, for example the XY model below the ordering or Berezinskii--Kosterlitz--Thouless (BKT)  temperature. Instead, the diagonal symmetry is preserved in the normal state that we are considering; this implies that $\phi_1 + \phi_2$ is not a NG field. 
The diffusive EFT should therefore contain $\phi_a$, but not $\phi_r$. Nevertheless, as reviewed in App.~\ref{app_review_Z2}, the simplest way to impose the collapse rules and KMS symmetry is to introduce a Keldysh partner for $\phi_a$: a field $\mu_r$, that has the same transformation properties as $\dot \phi_r$ would have. Note the key time derivative---$\mu_r$ is thus invariant under the $U(1)$ symmetries, and does not need to appear with derivatives in the action. The construction of the EFT is reviewed in App.~\ref{app_review_Z2}: to leading order in derivatives the action is 
\begin{equation}\label{eq_S2_Z2}
S = \chi \int_{t,x}i T D (\nabla\phi_a)^2 + \mu_r \left(\dot \phi_a + D \nabla^2 \phi_a\right)  + \cdots.
\end{equation}
where $\int_{t,x}\equiv \int dt d^d x$.
The physical density $ n \simeq \frac1{\chi}\mu_r$ satisfies the noisy diffusion equation to leading order in derivatives: $\dot n - D \nabla^2 n\simeq -i T \chi D\nabla^2 \phi_a$.

Let us now turn to the symmetry breaking pattern $U(1)^4 \to U(1)_{\rm diag}$ relevant for $Z_4$ and 4-point OTOCs. If the $U(1)^4$ symmetry were entirely spontaneously broken, the EFT would now contain four NG fields
\begin{equation}
\phi_1,\, \phi_2,\, \phi_3,\, \phi_4.\,
\end{equation}
In analogy with \eqref{eq_KeldyshRotation_Z2}, we introduce the basis:
\begin{equation}\label{eq_KeldyshRotation_Z4}
\begin{split}
\phi_R 
	&= \phi_1 + \phi_2 + \phi_3 + \phi_4\\
\phi_A
	&= \phi_1 - \phi_2 + \phi_3 - \phi_4\\
\phi_+
	&= \phi_1 + \phi_2 - \phi_3 - \phi_4\\
\phi_-
	&= \phi_1 - \phi_2 - \phi_3 + \phi_4\\
\end{split}
\end{equation}
The fields $\phi_A,\,\phi_+,\,\phi_-$ are NG fields that must enter with derivatives in the action. The fact that the diagonal symmetry is preserved implies that $\phi_R$ should not enter the EFT. However, the EFT will again contain a partner $\mu_R$, whose transformation properties under collapse or KMS are the same as $\dot\phi_R$. 

\subsection{Constraints on the EFT}

Having identified the relevant degrees of freedom, we would like to construct an effective field theory representation for the generating functional \eqref{eq_Z4}
\begin{equation}\label{eq_Z4_EFT}
Z_4[A]
	= \int D\mu_R D\phi_A D\phi_+ D \phi_- \, 
	e^{i S[\mu_R,\phi_A,\phi_+,\phi_-;A]}\, .
\end{equation}
The symmetry breaking pattern \eqref{eq_SSB} implies that the NG fields $\phi_A,\phi_+,\phi_-$ must enter the action with derivatives, while $\mu_R$ does not need derivatives.

The general constraints on $Z_4$ obtained in Sec.~\ref{ssec_Z4_properties} can be translated into constraints on the effective action $S$. While unitarity \eqref{eq_unitarity} and KMS \eqref{eq_KMS} essentially translate into symmetries of $S$, the collapse constraint \eqref{eq_collapse} is somewhat unusual in EFT: it does not relate the generating functional to itself upon a group action on the sources, and hence does not have the interpretation of a symmetry. Nevertheless, we show in App.~\ref{app_Z_to_S} that it leads to a simple condition on the action. In summary, the action must satisfy: 
\begin{enumerate}
\item
	{\em Collapse:} if any two neighboring fields are identified, the action is independent of that field and collapses to the 1-CTP EFT. In the $\phi_1,\phi_2,\phi_3,\phi_4$ basis this implies that $S[\phi_1,\phi_2,\phi_3,\phi_4]$ must satisfy:
	\begin{equation}\label{eq_action_collapse}
	\begin{split}
	&S[\phi_1,\phi_2,\phi,\phi]
		=S[\phi_1,\phi,\phi,\phi_2]\\
		&=S[\phi,\phi,\phi_1,\phi_2] = S_{\hbox{\scriptsize 1-CTP}}[\phi_1,\phi_2]\, .
	\end{split}
	\end{equation}
\item
	{\em Unitarity} imposes a reality condition on the action
	\begin{equation}\label{eq_action_unitarity}
	S^*[\phi_R,\phi_A,\phi_+,\phi_-] = -S[\phi_R,-\phi_A,-\phi_+,\phi_-]
	\end{equation}
\item
	{\em KMS} symmetry imposes
	\begin{equation}\label{eq_action_KMS}
	S[\phi] = S[\tilde \phi_{K^2}] = S[\tilde \phi_{TK}]\, .
	\end{equation}
\end{enumerate}
The KMS-transformed fields $\tilde \phi_{K^2},\tilde \phi_{TK}$ are given by
\begin{subequations}\label{eq_2CTP_KMS}
\begin{align}
	(\tilde \phi_{K^2})^I(\omega)&= (h_{K^2})^I{}_J(\omega) \phi^J(\omega)\\
	(\tilde \phi_{TK})^I(\omega)&= -(h_{TK})^I{}_J(\omega) \phi^J(-\omega)\, ,
\end{align}
\end{subequations}
with, in the 1234 basis,
\begin{align}
h_{K^2}
	&= 
\left(\begin{smallmatrix}
	&&1\\
	&&&1\\
	e^{\beta\omega}\\
	&e^{\beta\omega}
\end{smallmatrix}\right),&
h_{TK}
	&= \left(\begin{smallmatrix}
	e^{-\beta\omega}\\
	&&&1\\
	&&1\\
	&1
\end{smallmatrix}\right).
\end{align}
(The notation $K^n$ reflects the fact that we permuted legs across the density matrix $n$ times in Eq.~\eqref{eq_KMS}, and $T$ refers to time-reversal.)
In practice, it is often useful to implement KMS perturbatively in $\beta\omega$, see App.~\ref{app_terms}.


\subsection{Construction of the EFT}\label{ssec_construct_EFT}

We are now ready to construct the EFT for OTOCs. Consider first the Gaussian action
\begin{equation}
S_{(2)} = \frac12 \int_{p}M_{IJ}(p) \phi^I_{-p}\phi^J_{p} \,,
\end{equation}
with $p=\{\omega,\vec k\}$ and $\int_p \equiv \int \frac{d\omega d^dk}{(2\pi)^{d+1}}$. Imposing all the collapse constraints \eqref{eq_action_collapse} leads to a matrix in the ($RA+-$) basis
\begin{equation}
M_{IJ}(-p,p) = 
\left(\begin{array}{cccc}
0&b_p&0&0\\
b_{-p}&a_p&0&b_{-p}-b_p\\
0&0&0&b_p\\
0&b_p-b_{-p}&b_{-p}&0
\end{array}\right)\, .
\end{equation}
Imposing KMS \eqref{eq_action_KMS} then fixes
\begin{equation}
a(p) = 2 \coth \left(\frac{\beta \omega}{2}\right) (b(-p)-b(p))\, .
\end{equation}
Finally, imposing unitarity \eqref{eq_action_unitarity} leads to the relation
\begin{equation}\label{eq_b_unitarity}
b(p)^* = b(-p)\, , 
\end{equation}
which implies that $b$ is a real function (real Taylor coefficients) of $ip$. It also implies that $a(p)$ is purely imaginary (as a function of $p$), and even under $p\to -p$:
\begin{equation}\label{eq_a_of_b}
a(p)
	=-4i\coth \left(\frac{\beta \omega}{2}\right) \Im b(p)\, .
\end{equation}
We can now expand the action in $p$ (or derivatives). SWSSB implies that $\phi_R$ must appear with a time derivative $\mu_R\equiv \dot \phi_R$, and hence $b(p) \propto \omega$. Further using \eqref{eq_b_unitarity}, its small momentum expansion must take the form
\begin{equation}
b(p)
	= \frac\chi4 (-i\omega) \left(i\omega + D k^2+ \cdots\right) \, , 
\end{equation}
with $\chi,\,D\in \mathbb R$. Eq.~\eqref{eq_a_of_b} then implies
\begin{equation}
a(p)
	=  2iT\chi D k^2+ \cdots\, .
\end{equation}
The fact that the parameters $\chi,\,D$ correspond to the familiar charge susceptibility and diffusivity follows from \eqref{eq_action_collapse}, which requires the 2-CTP action under certain conditions to reduce to the 1-CTP one (regular fluctuating hydrodynamics). 

The final action in position space takes the form:
\begin{equation}\label{eq_S2_Z4}
\begin{split}
S_{(2)} &= 
	\frac{\chi}{4}\int_{tx} 4iT D (\nabla \phi^A)^2
	+ \mu^R (\dot \phi^A + D \nabla^2\phi^A) \\
	&\quad \ + \dot \phi^+ (\dot \phi^- + D \nabla^2\phi^-) 
	+ 2 D \nabla \dot \phi^A \nabla\phi^- + \cdots,
\end{split}
\end{equation}
where $\cdots$ denotes higher derivative terms. The only coefficients entering are the familiar thermodynamic and transport parameters $\chi$ and $D$ (or the conductivity $\sigma=\chi D$), which already appeared in the 1-CTP EFT. This is no surprise: the Gaussian action only captures two-point functions, which could have been obtained from the 1-CTP EFT. 

Let us now turn to interactions. In this paper, we focus on cubic and quartic terms:
\begin{align}
S_{(3)}
	&= \frac{1}{3!} \int_{p_1p_2p_3} \!\delta_{\Sigma_i p_i} M_{IJK}(\{p\}) \phi_{p_1}^I \phi_{p_2}^J \phi_{p_3}^K\, , \\
S_{(4)}
	&= \frac{1}{4!} \int_{p_1p_2p_3p_4} \!\delta_{\Sigma_i p_i} M_{IJKL}(\{p\}) \phi_{p_1}^I \phi_{p_2}^J \phi_{p_3}^K\phi_{p_4}^L\, , 
\end{align}
with $\delta_p \equiv (2\pi)^{d+1}\delta(\omega)\delta^d(\vec k)$. These will give the leading contributions to the simplest OTOCs, which are 4-point functions. 

We impose the constraints on the cubic and quartic action in App.~\ref{app_terms} to find the leading order terms. For the cubic action, the only terms we find are equivalent to cubic terms entering the 1-CTP EFT \cite{Delacretaz:2023ypv} (the leading order ones are related to derivatives of thermodynamic or transport parameters with respect to chemical potential: $\chi'\equiv d \chi/d\mu$ and $\sigma' \equiv d\sigma/d\mu$). Importantly, there is no parameter in the cubic 2-CTP action that does not appear in the 1-CTP action. This is to be expected, since these parameters contribute to 3-point functions, and any 3-point function in a thermal state can be represented on a 1-CTP contour (this would not be the case for a non-thermal density matrix---see, e.g., \cite{Chaudhuri:2018ihk}). 

The quartic action instead can feature new terms not present in the 1-CTP EFT, because there are 4-point OTOCs that cannot be represented on a 1-CTP. The leading 1-CTP quartic terms were studied in \cite{Delacretaz:2023ypv} and are reviewed in App.~\ref{app_terms}: they again have the interpretation of derivatives of familiar transport parameters $\chi'',\,\sigma''$. These terms of course have incarnations in the  2-CTP EFT. At leading order in derivatives and scaling, we find two new terms that drop out of any 1-CTP correlation function:
\begin{equation}\label{eq_lambda12}
S_{(4)} = \int dt d^dx \, \lambda_1 \mathcal L_1 +\lambda_2 \mathcal L_2 
\end{equation}
with
\begin{align}
\mathcal L_1
	&=i\left[\dot{\phi}_-^2 \left(\nabla{\phi}_+^2- \nabla{\phi}_A^2\right)+\nabla{\phi}_-^2 \left(\dot{\phi}_+^2-\dot{\phi}_A^2\right)+\cdots \right]\notag \\ \notag 
\mathcal L_2
	&= i\left[\dot{\phi}_- \nabla{\phi}_- (\dot{\phi}_+ \nabla{\phi}_+-\dot{\phi}_A \nabla{\phi}_A)+\cdots \right]
\end{align}
The unitarity constraint \eqref{eq_action_unitarity} implies that these coefficients are real $\lambda_{1,2}\in \mathbb R$. The ellipses denote terms that are higher order in scaling (see Sec.~\ref{ssec_scaling_EFT}), as well as terms $\propto \beta$ that are imposed by KMS symmetry; these are spelled out in Eqs.~\eqref{eq_app_l1} and \eqref{eq_app_l2}. \new{These coefficients also enter in the constitutive relation for the current $j_i^R \equiv 4\delta S / \delta A_i^A$:
\begin{equation}
    \begin{split}
  j_{i,R}&=\sigma\left(E_{i,R}-\partial_i\mu_R- 2 \partial_i \dot\phi_- + 8i T \partial_i \phi_A\right) +\cdots\\
    &+4i\left(\lambda_1 \dot{\phi}_-^2\partial_i\phi_A+\lambda_2\dot{\phi}_-\dot{\phi}_A\partial_i\phi_-+\cdots\right)  \, .   
    \end{split}
\end{equation}    
The first line is the usual constitutive relation for the current of a diffusing charge, albeit with an unusual basis of noise fields due to the 2-CTP representation.
The new out-of-time ordered transport parameters $\lambda_{1,2}$ can be interpreted as a form of nonlinear response involving correlated noise between the two replica.
}

\section{OTOCs and their transport parameters}\label{sec_EFT_properties}

Our construction revealed novel transport parameters for diffusive systems that are only visible in OTOCs. We explore these parameters in this section, and use the EFT to predict the behavior of OTOCs more generally.

\subsection{Scaling in the EFT}\label{ssec_scaling_EFT}

The control parameter underlying the validity of the EFT is small frequency $\omega$ or wavevector $q$. Corrections to the Gaussian action \eqref{eq_S2_Z2} or \eqref{eq_S2_Z4} can be classified according to their scaling dimension to anticipate their contributions to observables. The interesting scaling regime for diffusion is $\omega\sim q^2$; in this regime, the 1-CTP EFT \eqref{eq_S2_Z2} describing conventional fluctuating diffusion shows that the fields scale as
\begin{equation}
\mu^r \sim \phi^a \sim q^{d/2}\, .
\end{equation}
For regular densities $n\sim \mu^r$, this scaling indeed predicts that diffusive correlators $\langle nn\rangle\sim 1/t^{d/2}$. Applying the same logic to the 2-CTP Gaussian action \eqref{eq_S2_Z4}, one finds
\begin{equation}\label{eq_2CTP_scaling}
\mu^R \sim \phi^A \sim \dot \phi^- \sim \phi^+ \sim q^{d/2}\, .
\end{equation}
Note that one of the Goldstones ($\phi^-$) is more relevant. The advantage of the Keldysh basis introduced in \eqref{eq_KeldyshRotation_Z4} is precisely that it is an eigenbasis for scaling. This scaling underpins the construction of the EFT in Sec.~\ref{ssec_construct_EFT}.

In the 1-CTP action, the leading quartic terms $\chi''$ and $\sigma''$ have two extra powers of $\mu^r$, so that 
\begin{equation}
\frac{S^{(4)}_{\rm toc}}{S^{(2)}}\sim (\mu^r)^2 \sim q^d\, .
\end{equation}
The new OTOC parameters found in \eqref{eq_lambda12} have the same scaling 
\begin{equation}\label{eq_vertex_scaling}
\frac{S^{(4)}_{\rm otoc}}{S^{(2)}}\sim q^d\, .
\end{equation}
In particular, corrections to diffusion are irrelevant, and the dynamics is controlled to leading order by the Gaussian diffusive fixed point. This already justifies part of the heuristic argument in Sec.~\ref{sec_heuristic}: higher-point TOCs and OTOCs factorize into products of two-point functions (said differently, any operator overlapping with hydrodynamic fields approximately satisfies Wick's theorem).

While the dynamics is described to leading order by a simple diffusive fixed point,  many observables vanish at the fixed point---this is the case for the connected OTOCs of interest in this paper. These observables are thus controlled by the leading irrelevant corrections to the diffusive fixed points, namely $\chi'',\,\sigma''$, and the OTO-transport parameters $\lambda_{1,2}$.

\subsection{Predicting the scaling of observables}\label{ssec_scaling_obs}

We are now ready to make predictions for OTOCs in generic thermalizing quantum many-body systems using the EFT. A simple argument shows that there are only three independent 4-point functions of identical operators $O$ in a thermal state $\rho = e^{-\beta H}/\Tr e^{-\beta H}$ \cite{Wang:1998wg,Haehl:2017eob}. Let $t_1<t_2<t_3<t_4$. Using trace cyclicity, commuting past the thermal density matrix at the cost of a shift in imaginary time, one can make $O(t_1)$ the right-most operator. Every 4-point function can then be related to the correlators
\begin{subequations}\label{eq_4pt_basis}
\begin{align}
g_0
	&= \Tr \left(\rho O(t_4)O(t_3)O(t_2)O(t_1)\right)\\
g_1
	&= \Tr \left(\rho O(t_3)O(t_4)O(t_2)O(t_1)\right)\\
g_2
	&= \Tr \left(\rho O(t_4)O(t_2)O(t_3)O(t_1)\right)\, ,
\end{align}
\end{subequations}
and their complex conjugates. $g_0$ is the regular time-ordered Wightman function. $g_1$ is not time-ordered, but can be represented on a 1-CTP. Finally, $g_2$ is a genuine OTOC, that requires a 2-CTP.

The 2-CTP Keldysh basis of correlators 
\begin{equation}
G_{IJKL} \equiv \langle O_I(t_1) O_J(t_2) O_K(t_3) O_L(t_4)\rangle\, , 
\end{equation}
with $I,J,K,L \in \{1,2,3,4\}$, is highly redundant; each can be expressed in terms of the three complex functions in \eqref{eq_4pt_basis}. We provide the map between these two bases in App.~\ref{app_scaling_corr}. 

Having expressed physical observables in terms of the Keldysh fields with simple scaling properties, it is now straightforward to estimate the observables \eqref{eq_4pt_basis}. As discussed above, they approximately factorize into a disconnected piece; we then focus on the connected versions $g_i^c$ of these correlators, $i=0,1,2$. We find that they are all approximately equal, and given by the largest Keldysh correlator
\begin{equation}
g_0^c,g_1^c,g_2^c \sim G_{RRRR} \sim \frac{1}{t^{3d/2}}\, .
\end{equation}
The last relation follows from $(\mu^R)^{4}\sim q^{2d}\sim 1/t^d$ using \eqref{eq_2CTP_scaling}, and the fact that one vertex \eqref{eq_vertex_scaling} costs $q^{d}\sim 1/t^{d/2}$. The fact that $g_0^c\simeq g_1^c \simeq g_2^c$ at late times is the EFT derivation for the heuristic argument of Sec.~\ref{sec_heuristic}: operators approximately commute at late times. As discussed there, it motivates studying the difference between the $g_i^c$, which involve commutators. This is carried out in detail in App.~\ref{app_scaling_corr}, with results reported in Table \ref{tab:scaling_correlators}.

\begin{table}[h]
    \centering
    \begin{tabular}{|l|cc|}
        \hline
        & $\beta \neq 0$ & $\beta = 0$ \\
        \hline
        $\mathrm{Re}(g_0)$          & $t^{-3d/2}$ & $t^{-3d/2}$ \\
        $\mathrm{Im}(g_0)$          & $t^{-(3d/2+1)}$ & $t^{-(3d/2+2)}$ \\
        $\mathrm{Re}(g_0 - g_1)$    & $t^{-(3d/2+2)}$ & $t^{-(3d/2+3)}$ \\
        $\mathrm{Im}(g_0 - g_1)$    & $t^{-(3d/2+1)}$ & $t^{-(3d/2+2)}$ \\
        $\mathrm{Re}(g_0 - g_2)$    & $t^{-(3d/2+2)}$ & $t^{-(3d/2+2)}$ \\
        $\mathrm{Im}(g_0 - g_2)$    & $t^{-(3d/2+1)}$ & $t^{-(3d/2+2)}$ \\
        \hline
    \end{tabular}
    \caption{EFT prediction for the scaling of the connected correlators defined in Eq.~\eqref{eq_4pt_basis}. While the first column directly follows from the scaling analysis, several cancellations at $\beta=0$ lead to smaller correlators. }
    \label{tab:scaling_correlators}
\end{table}

\subsection{Kubo formulas for OTOC transport parameters}

Beyond the scaling behavior of observables, the EFT provides predictions for the entire spatial and time resolved correlators
\begin{figure}[t]
\centering

\begin{tikzpicture}[scale=0.8]
  \coordinate (L) at (-0.9,0);
  \coordinate (R) at ( 0.9,0);

  \coordinate (L1) at (-1.9,  1.1);
  \coordinate (L2) at (-1.9, -1.1);
  \coordinate (R1) at ( 1.9,  1.1);
  \coordinate (R2) at ( 1.9, -1.1);

  \draw[thick] (L) -- (L1);
  \draw[thick] (L) -- (L2);
  \draw[thick] (R) -- (R1);
  \draw[thick] (R) -- (R2);

  \draw[thick] (L) -- (R);

  \fill (L) circle (1.5pt);
  \fill (R) circle (1.5pt);
\end{tikzpicture}
\hspace{2.0cm}
\begin{tikzpicture}[scale=0.8]
  \coordinate (C) at (0,0);

  \coordinate (A) at ( 1.2,  1.2);
  \coordinate (B) at (-1.2,  1.2);
  \coordinate (D) at ( 1.2, -1.2);
  \coordinate (E) at (-1.2, -1.2);

  \draw[thick] (C) -- (A);
  \draw[thick] (C) -- (B);
  \draw[thick] (C) -- (D);
  \draw[thick] (C) -- (E);

  \fill (C) circle (1.5pt);
\end{tikzpicture}

\caption{Feynman diagrams that contribute to OTOCs in the EFT. Cubic vertices are forbidden in the presence of particle-hole symmetry, in which case only the right diagram exists. There are only four different quartic vertices at leading order: two which can be measured in regular transport ($\chi''$ and $\sigma''$), and the two OTO-transport parameters $\lambda_1$ and $\lambda_2$.}
\label{fig_Gmmpp_feyndiagram}
\end{figure}
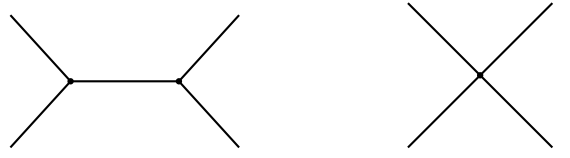

\begin{equation}
\Tr \left( \rho n(t_4,x_4)n(t_3,x_3)n(t_2,x_2)n(t_1,x_1)\right)\, .
\end{equation}
As in Eq.~\eqref{eq_4pt_basis}, these can be separated into correlators that are representable on a 1-CTP ($g_0$ and $g_1$), and a correlator that requires a 2-CTP ($g_2$). The former were studied in \cite{Delacretaz:2023ypv} and are briefly reviewed in App.~\ref{app_review_Z2}. Here we focus on the genuine OTOC, $g_2$, which probes the novel OTO-transport parameters $\lambda_{1,2}$. As discussed above, $g_2$ is approximately equal to the time-ordered correlator $g_0$ at late times, and the interesting observable is $g_2 - g_0$, as anticipated in Eq.~\eqref{eq_otoc_commutator}. We show in App.~\ref{app_scaling_corr} that its real and imaginary parts can be obtained with the following Keldysh correlators: 
\begin{equation}
\begin{split}
\Re(g_2 - g_0) &\simeq \frac1{16}G_{--++} \, , \\
\Im(g_2 - g_0)&\simeq-\frac1{32}G_{R-+R}  \, .
\end{split}
\end{equation}
%
(These expressions hold at $\beta=0$, the generalization to $\beta\neq 0$ are shown in App.~\ref{app_scaling_corr}.) The leading behavior of these correlators is captured by a tree-level diagram in the EFT, shown in Fig.~\ref{fig_Gmmpp_feyndiagram}. Their full expression in momentum space is given in Eq.~\eqref{eq_app_Gmmpp}. Remarkably, the real and imaginary parts of this OTOC have qualitatively different behavior: \new{at leading order in the late time expansion}, the imaginary part is entirely determined by conventional transport data that can be extracted from time-ordered correlators (specifically, $\chi$ and $\sigma$, and their derivatives with respect to thermodynamic parameters), whereas the real part reveals the OTO-transport parameters identified in Eq.~\eqref{eq_lambda12}. As with conventional transport parameters, limits of these correlators can be used to establish Kubo formulas for the OTO-transport parameters $\lambda_1,\,\lambda_2$  identified in \eqref{eq_lambda12}. In particular, the static limit $\omega_i\to 0$ of $G_{--++}$ involves both $\lambda_1$ and $\lambda_2$, see Eq.~\eqref{eq_app_Kubo_freq}. Different kinematic configurations of the wavevectors $k_i$ can then be used to isolate each parameter separately, for example:
\begin{equation}\label{eq_Kubo}
\begin{split}
&\lim_{\omega_{i}\to 0}\lim_{\substack{k_4\to -k_3,\\ k_2\to -k_1}}G_{--++}(p_1,p_2,p_3,p_4) = -\frac{4\lambda_1}{\beta D^2}\frac{(k_1^2+k_3^2)}{k_1^2k_3^2}\\
&\lim_{\omega_{i}\to 0}\lim_{\substack{k_1,k_2\to k,\\k_3,k_4\to -k}}G_{--++}(p_1,p_2,p_3,p_4)=\frac{4}{\beta D^2}\frac{(2\lambda_1+\lambda_2)}{k^2}.
\end{split}
\end{equation}
Here, we have assumed particle-hole symmetry for simplicity to forbid the cubic EFT terms $\chi',\sigma'$, which would otherwise give additional contributions to \eqref{eq_Kubo}. These Kubo formulas could help identify OTOC parameters in tractable microscopic models.

It is of course also possible to measure the OTO-transport parameters $\lambda_{1,2}$ in time domain. For example, the coefficient of the $1/t^{3d/2 + 2}$ tail in $\Re (g_0 - g_2)$ probes a linear combination of $\lambda_1$ and $\lambda_2$, and is measured numerically in $d=1$ models in Sec.~\ref{sec_numerics}.



\subsection{Bounds on OTOC transport parameters}

In addition to the conditions (\ref{eq_collapse}--\ref{eq_KMS}), the generating functional $Z[\{A\}]$ also satisfies a bound 
\begin{equation}
|Z| \leq 1\, , 
\end{equation}
which follows from unitarity \cite{Glorioso:2016gsa,Jensen:2018hse,Liu:2018kfw}. This bound is closely related to the positivity of spectral densities in unitary quantum mechanics. It is often assumed that this condition translates into the constraint on the hydrodynamic effective action
\begin{equation}\label{eq_ImSpos}
\Im S \geq 0\, , 
\end{equation}
which must hold for any configuration of the fields (in our case, $\mu_R,\,\phi_A,\,\phi_-,\,\phi_+$). To leading order in fields and derivatives, applying this bound to the action \eqref{eq_S2_Z2} or \eqref{eq_S2_Z4} correctly requires the conductivity to be positive: $\sigma = \chi D\geq 0$. However, we show in App.~\ref{app_Z_to_S} that this condition is too strong, and leads to incorrect conclusions in general. Here, we propose a slightly different criterion, and explore its consequences on OTOC parameters.

In the Gaussian action \eqref{eq_S2_Z4}, notice that only field configurations involving $\nabla\phi_A\neq 0$ dissipate. To obtain nontrivial positivity bounds beyond $\sigma \geq 0$, \eqref{eq_ImSpos} would suggest to consider configurations involving the other fields, or $\phi_A = \phi_A(t)$. As discussed in App.~\ref{app_Z_to_S}, this leads to unphysical constraints. We instead consider field configurations with $\phi_A = 0$. For the conventional 1-CTP EFT, this would reduce the entire action to 0, and no positivity bound other than $\sigma \geq 0$ is obtained (as expected). Interestingly, the 2-CTP EFT instead has an imaginary part that does not involve $\phi_A$, coming entirely from the OTO-transport parameters. This imaginary part from Eq.~\eqref{eq_lambda12} is 
\begin{equation}
\begin{split}
\Im \mathcal L_{(4)}[\phi_A = 0]
	&= \lambda_1 \left[(\dot \phi_+\nabla\phi_-)^2 + (\dot \phi_-\nabla\phi_+)^2\right]\\
	&+ \lambda_2\, \dot \phi_+\dot \phi_- (\nabla\phi_-\cdot \nabla\phi_-)\, .
\end{split}
\end{equation}
This has the form $\lambda_1 (a^2 + b^2) + \lambda_2 a\cdot b$, and thus is non-negative if the OTO-transport parameters satisfy
\begin{equation}
\lambda_1 \geq \frac12 |\lambda_2| > 0\, .
\end{equation}
It is surprising for the coefficients of irrelevant operators to be bounded by positivity of dissipation in hydrodynamics---indeed this does not occur in the conventional 1-CTP EFT---because dissipation is usually dominated by the leading transport parameters. We emphasize that the argument above is not rigorous; it would be interesting to prove these conjectured bounds, perhaps by considering spectral densities of bilocal operators $n(t,x)n(t',x')$ that are dominated by the OTO-transport parameters. We leave this for future work.

\subsection{Regime of validity of the EFT}\label{ssec_regime}

\begin{figure}[t]
    \centering
    \begin{tikzpicture}[scale=0.8, >=stealth]

      \draw[->, very thin] (-3.5,0) -- (3.5,0) node[anchor=west] {$x$};
      \draw[->, very thin] (0,0) -- (0,4.2) node[anchor=south] {$t$};

      \draw[thick, draw=gray!60!black] (-3.0,4.0) -- (0,0) -- (3.0,4.0);

      \draw[thick, dashed, draw=blue, domain=-2:2, samples=200]
        plot (\x, {\x*\x});

      \fill (0,1.5) circle (1.2pt);
      \node[left] at (0,1.5) {$\tau_{\mathrm{eq}}$};

      \node[anchor=west] at (1.8,2.3) {$t = |x|/v$};
      \node[anchor=west] at (0.3,4.2) {\color{blue}$t = x^{z>1}$};

    \end{tikzpicture}

    \caption{Regime of validity of the EFT versus the butterfly cone. The EFT requires $t\gtrsim \tau_{\rm eq}$, and $t\gtrsim x^{z}$ for $z>1$.}
    \label{fig_hydro-regime}
\end{figure}
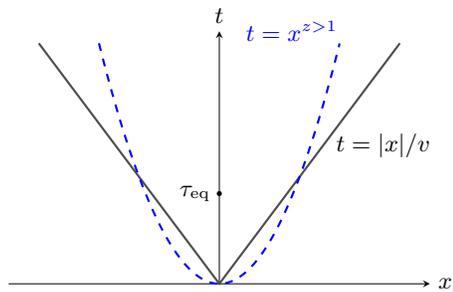

The effective field theory constructed in Sec.~\ref{sec_EFT_construction} is an expansion in derivatives and fluctuations, as is conventional in fluctuating hydrodynamics. This, together with the observation that fluctuation corrections are irrelevant (Eq.~\eqref{eq_vertex_scaling}), guarantees that the EFT provides a controlled expansion for small frequencies and wavevectors $\omega , \, Dk^2 \ll 1/\tau_{\rm eq}$. However, interestingly, this does not imply that it is controlled at late times $t\gg \tau_{\rm eq}$ for any $x$. This is simplest to see by considering gradient corrections to the diffusion equation:
\begin{equation}
\dot n + \nabla\cdot j=0\,, \quad 
	j = - D(-\nabla^2)\nabla n + \cdots\, ,
\end{equation}
with $D(k^2) = D_0 + D_1 k^2 + D_2 k^4 + \cdots$\, .
The higher-derivative terms $D_{\geq 1}$ will clearly give small corrections to momentum space observables at small wavevector $k$. For example, the symmetric Green's function for $\omega\ll T$ is 
\begin{equation}
G(\omega,k) = \frac{2T \chi(k^2) D(k^2)}{\omega^2 + [D(k^2)k^2]^2}\, ,
\end{equation}
where we are also allowing for higher-derivative corrections to the static susceptibility $\chi(k^2) = \chi_0 + \chi_1 k^2 + \cdots$. The Fourier transform can be expressed in terms of a differential operator acting on the usual diffusive correlator:
\begin{equation}\label{eq_Gtx_exact}
G(t,x)
	= T\chi(-\nabla^2) e^{\delta D(-\nabla^2)\nabla^2 t} \frac{1}{\sqrt{4\pi D t}} e^{-x^2/(4Dt)}\, ,
\end{equation}
with $\delta D(k^2) = D(k^2) - D_0$. Evaluating this in the diffusive region $x\sim \sqrt{D_0 t}$, the gradient  expansion becomes an expansion in $1/t$ and is thus controlled at late times
\begin{equation}
G(t,x) \sim 
	\left[\chi_0 + \frac{\chi_1 }{D_0 t} + \cdots \right]
	\left[D_0 + \frac{D_1 }{D_0 t} + \cdots \right]e^{-x^2/(4Dt)}.
\end{equation}
However, evaluating the correlator along a sound front $x = v t$, the expansion is uncontrolled:
\begin{equation}
\begin{split}
G(t,x) \sim
	&\left[\chi_0 + \chi_1 \frac{v^2}{D_0^2} + \cdots \right]\\
	&\times \left[D_0 + D_1 \frac{v^4}{D_0^3}t + \cdots \right]e^{-x^2/(4Dt)}\, .
\end{split}
\end{equation}
The higher-derivative corrections in $\chi(k^2)$ are already problematic as they give $O(1)$ corrections in this region. The corrections in $D(k^2)$ are worse: they grow! One can check that control is recovered along any trajectory $x\sim t^{1/z_{\rm traj}}$ with $1<z_{\rm traj}$, although if $z_{\rm traj}<2$ one needs to keep derivative corrections $D_n$ up to $n = \lfloor\frac{2-z_{\rm traj}}{2(1-z_{\rm traj})}\rfloor$ to capture the {\em leading} late time behavior of the correlator.

The upshot of this discussion is that predictions from the diffusive EFT for spacetime correlators are controlled in a broad `cone' that slightly bends inward, and does not include any sound-cone at late times. This is notable, because for certain models one expects `butterfly cones' in OTOCs \cite{Khemani:2017nda,xu2022scrambling}. These are not at tension with the EFT results, which do not predict such a ballistically propagating excitation, precisely because the EFT becomes uncontrolled in this region. These non-overlapping regimes of validity as shown in Figure-\ref{fig_hydro-regime} make it challenging to establish an EFT that captures both diffusive fronts and butterfly cones in a controlled way. It would be interesting to revisit the results of \cite{Gao:2023wun,Choi:2023mab} from this perspective.

\subsection{Flow of time}

\begin{figure}
\centerline{
\hfill
\subfloat[]{

	\begin{tikzpicture}[thick,scale=0.7,
		fermion/.style={postaction={decorate}, decoration={markings,mark=at position 0.5 with {\arrow{>}}}}
	  ]

	  \coordinate (a) at (0,0);
	  \coordinate (r) at (3,0);
	  \coordinate (c) at (3.5,0); 
	  \coordinate (d) at ($2*(c)-(r)$); 

	  \fill (a) circle (2pt) node[above] {$a$};
	  \fill (r) circle (2pt) node[above] {$r\ \ $};

	  \draw[fermion] (a) -- (r);

	  \draw (c) circle (0.5);

	  \node at (d) {$=$};
	  
	  \coordinate (e) at (0.8,-1);
	  \node at (e) {};

	\end{tikzpicture}
} 
\hfill
\subfloat[]{\label{sfig_label2}

	\begin{tikzpicture}[thick,scale=0.7,
		fermion/.style={postaction={decorate}, decoration={markings,mark=at position 0.5 with {\arrow{>}}}}
	  ]

	  \coordinate (a) at (0,0);
	  \coordinate (r) at (3,0);
	  \coordinate (m) at (0,-1);
	  \coordinate (p) at (3,-1);
	  \coordinate (c) at (3.5,0); 
	  \coordinate (d) at ($2*(c)-(r)$); 

	  \fill (a) circle (2pt) node[above] {$A$};
	  \fill (r) circle (2pt) node[above] {$R\ \ $};
	  \fill (m) circle (2pt) node[below] {$-$};
	  \fill (p) circle (2pt) node[below] {$+\ \ $};

	  \draw[fermion] (a) -- (r);
	  \draw[fermion] (m) -- (p);
	  \draw (p) -- (r);

	  \draw (c) circle (0.5);

	  \node at (d) {$=$};
	  \node at ($.5*(p)+.5*(r)$) {$=$};

	\end{tikzpicture}
\hfill
}
}\caption{Flow of time in (a) 1-CTP and (b) 2-CTP Schwinger-Keldysh correlators. The left image illustrates the two non-zero correlators $G_{rr},\, G_{ra}$, and the fact that $G_{ra}(t) = 0$ for $t<0$. The right image shows the corresponding four nonzero 2-CTP propagators and their causal properties. \label{fig_flowoftime}}
\end{figure}
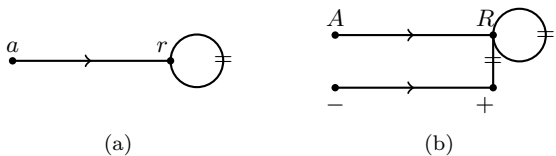

Real time Schwinger-Keldysh quantum field theories have an interesting causal structure that leads to a notion of `flow of time' in Feynman diagrams \cite{Caron-Huot:2008dyw,Gao:2018bxz}. This property comes from the latest time conditions, discussed in detail in App.~\ref{app_Z_to_S}. It leads to several diagrammatic simplifications that are not apparent from straightforward symmetry principles \cite{Caron-Huot:2008dyw,Gao:2018bxz,Michailidis:2023mkd}: otherwise valid Feynman diagrams that do not respect this flow of time structure will inevitably vanish upon evaluation. 

In the conventional 1-CTP EFT, $\phi_a$ must flow to the future to $\phi_r$, while $\langle \phi_r\phi_r\rangle$ is nonzero for any time ordering. A generalized flow of time is at play in our 2-CTP EFT: $\phi_A$ must flow to the future to $\phi_R$, and $\phi_-$ must flow to the future to $\phi_+$, while $\langle \phi_R\phi_R\rangle$ and $\langle \phi_R \phi_+ \rangle$ are nonzero for any time ordering. This causal structure is illustrated in Fig.~\ref{fig_flowoftime}. In terms of equations of motion, the fields $\phi_A$ and $\phi_-$ satisfy an advanced diffusion equation (like $\phi_a$), while $\phi_R\sim n$ and $\phi_+$ satisfy the regular diffusion equation (like $\phi_r$).

\section{Comparison to Numerics}\label{sec_numerics}
To test our analytical predictions, now we present numerical simulations performed in two one-dimensional spin models, which exhibit diffusive transport. As a first model, we consider the spin-$\frac{1}{2}$ mixed field Ising model
\begin{equation}
H=\sum_{n=1}^{L}J\sigma_{n}^{z}\sigma_{n+1}^{z}+g_{z}\sigma_{n}^{z}+g_{x}\sigma_{n}^{x} ,
\end{equation}
where $\sigma_n^\mu$ are Pauli operators in the $\mu=x, y, z$ direction and we use periodic boundary condition ${\sigma}_{L+1}^{\mu}={\sigma}_{1}^{\mu}$. Energy is the only conserved quantity in this model, the lattice energy density operator is
\begin{equation}
    h_{n}=J\sigma_{n}^{z}\sigma_{n+1}^{z}+\frac{1}{2}g_{z}(\sigma_{n}^{z}+\sigma_{n+1}^{z})+\frac{1}{2}g_{x}(\sigma_{n}^{x}+\sigma_{n+1}^{x}) .
\end{equation}
The parameters considered here are $J = 1.0$, $g_z=0.9$, $g_x=1.1$.

\begin{figure}[t]
	\includegraphics[width=0.9\columnwidth]{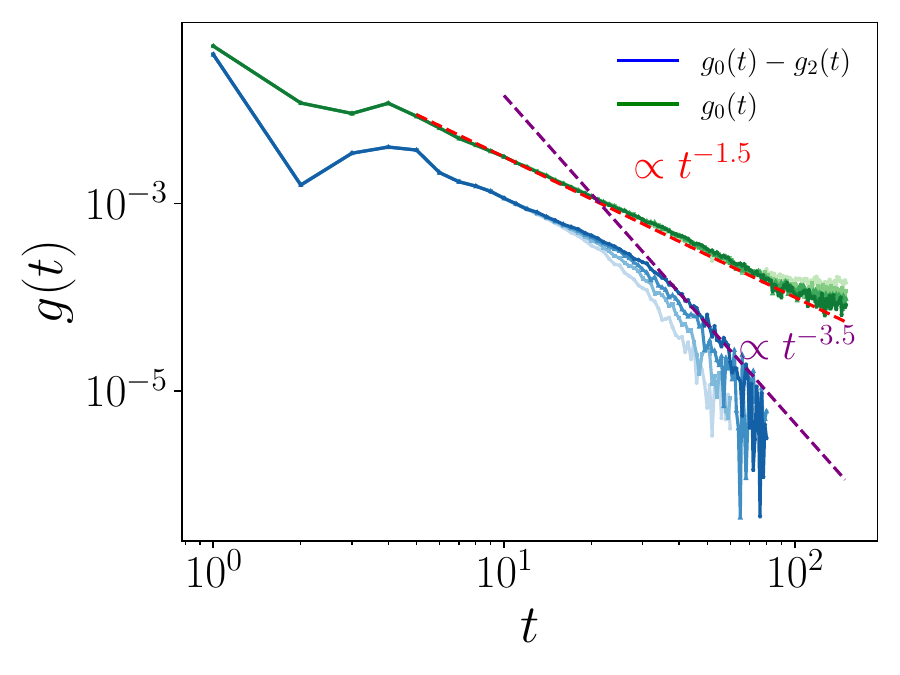}
\caption{Measurement of the OTO-transport parameters in the Floquet XXZ model for $\mu = 0.0$ and parameters ${\cal J} = \frac{\pi}{8},\ {\cal J}^\prime = \frac{\pi}{4},\ \lambda = \new{-\frac{\pi}{8}}$. The dashed lines indicate EFT prediction in Table~\ref{tab:scaling_correlators}. 
 Here the system sizes are $L=24,26,28,30$ (from light to dark). 
}
\label{fig_num_lambda12}
\end{figure}

As a second model, we consider a Floquet XXZ model with Floquet operator given by
\new{
\begin{equation}
{\cal U}={\cal U}_{\text{odd}}{\cal U}_{\text{even}}{\cal U}_{\text{nn}}.
\end{equation}}
Here
\begin{equation}
    {\cal U}_{\text{odd}}=\prod_{n=1}^{N/2}U_{2n,2n+1}\ ,\ {\cal U}_{\text{even}}=\prod_{n=1}^{N/2}U_{2n-1,2n},
\end{equation}
where
\begin{equation}
    U_{n,n+1}=e^{-i{\cal J}(\sigma_{n}^{x}\sigma_{n+1}^{x}+\sigma_{n}^{y}\sigma_{n+1}^{y})-i{\cal J}^{\prime}(\sigma_{n}^{z}\sigma_{n+1}^{z}-1)},
\end{equation}
and
\begin{equation}
    {\cal U}_{\text{nn}}=e^{-i\lambda\sum_{n=1}^{N}\sigma_{n}^{z}\sigma_{n+2}^{z}} .
\end{equation}
The total magnetization along the $z$ direction
\begin{equation}
    S^{\text{tot}}_z = \sum_{n=1} ^{N} \sigma^z_n
\end{equation}
is the only conserved quantity. This system has no energy conservation and equilibrates to a maximally mixed state (corresponding to $\beta=0$). We consider states of the form $\rho=e^{-\mu S^{\text{tot}}_z}$ with different values of $\mu$. 
When $\lambda = 0$ (i.e., in the absence of next-nearest-neighbor coupling), the system is integrable and exhibits various types of spin transport---ballistic, diffusive, or anomalous---depending on the values of ${\cal J}$ and ${\cal J}^\prime$. For $\lambda \neq 0$, the system becomes non-integrable, and the transport behavior is generally expected to be diffusive \cite{matthies2025thermalizations}. Throughout the numerical simulations, we consider two different parameters ${\cal J} = \frac{\pi}{8},\ {\cal J}^\prime = \frac{\pi}{4},\ \lambda = \frac{\pi}{8}$ and ${\cal J} = \frac{\pi}{10},\ {\cal J}^\prime = \frac{\pi}{4},\ \lambda = \frac{\pi}{10}$.
We consider again periodic boundary conditions. Our numerical method is based on dynamical quantum typicality \cite{PhysRevLett.102.110403-dqt-gemmer, HeitmannRichterSchubertSteinigeweg-DQT}, an approach that has been widely applied to quantum transport \cite{PhysRevLett.112.120601-dqt-robin,PhysRevB.95.035155-DQT,PhysRevE.108.024102-DQT,Robin_diffusion_2015-DQT,Robin_diffusion_2017-DQT} and recently extended to the study of OTOCs \cite{PhysRevResearch.2.043047-otoc-dqt} (see also \cite{Luitz:2017jrn}). 
In contrast to two-point correlation functions, the computational complexity of OTOC simulations scales as $\mathcal{O}(t^2)$, rendering the simulations substantially more challenging.

\begin{figure}[t]
\includegraphics[width=1\columnwidth]{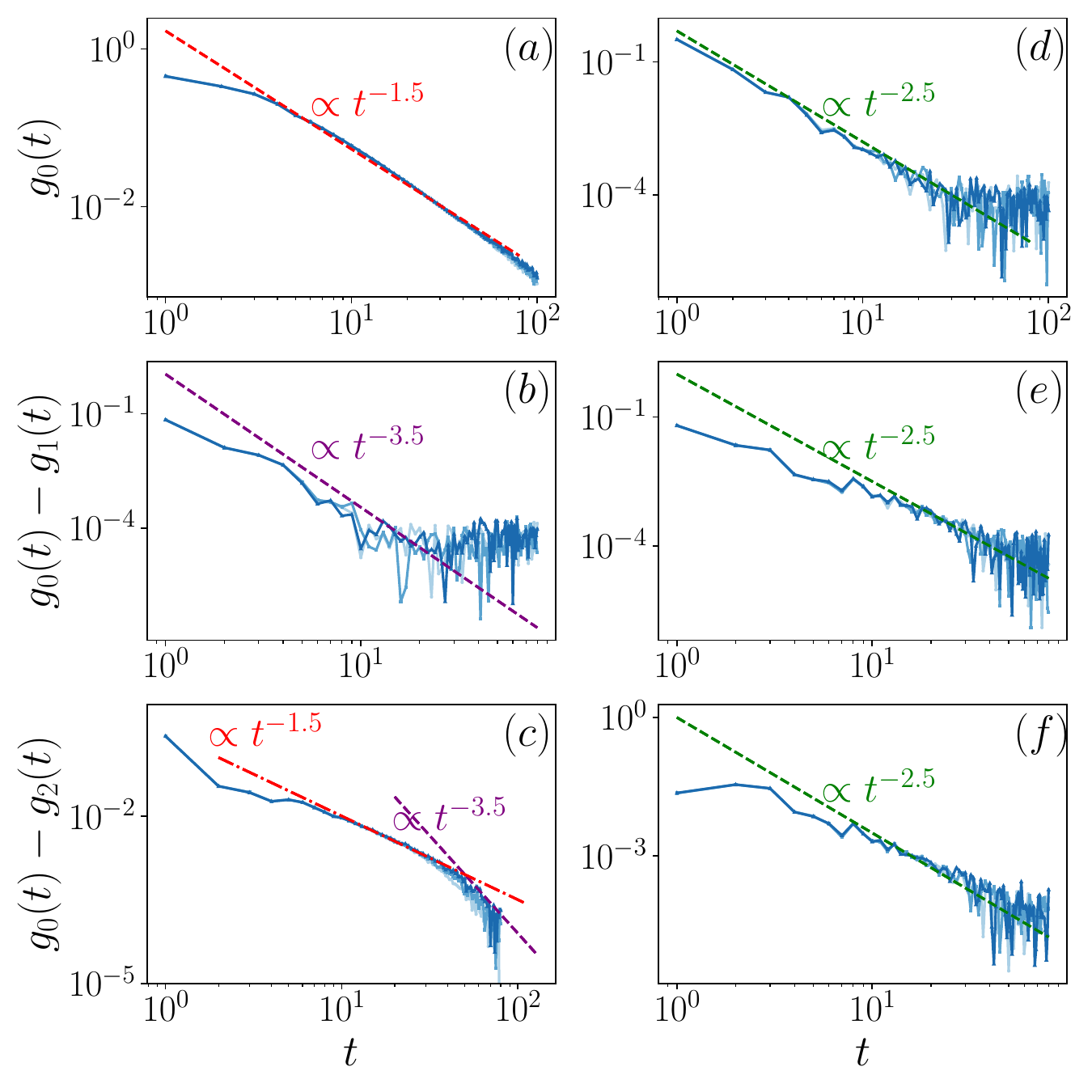}
\caption{Results in the mixed-field Ising model at finite temperature $\beta = 0.2$:
$g_0(t)$ [(a)(d)]; $g_0(t)-g_1(t)$ [(b)(e)] and $g_0(t)-g_2(t)$ [(c)(f)].
The real and imaginary parts are presented in the left and right panels, respectively.
The dashed lines indicate EFT prediction in Table~\ref{tab:scaling_correlators}. The dash--dot line in panel (c) is shown as a guide to the eye. Here the system sizes are $L=24,26,28$ (from light to dark).
}
\label{Fig-Ising-finiteT}
\end{figure}

We first reproduce the findings from Refs.~\cite{Khemani:2017nda,Rakovszky:2017qit} for the $1/\sqrt{t}$ scaling of OTOCs of operators at coincident times, see Fig.~\ref{fig_powerlaws}. As explained in Sec.~\ref{sec_heuristic}, this scaling arises in the EFT from an OPE of the coincident operators, which reduces the OTOC to a two-point function. Fig.~\ref{fig_powerlaws} further shows that the OTOCs at non-coincident points also have a behavior in agreement with the EFT predictions: a $1/t$ disconnected part, and a $1/t^{3/2}$ connected part.

As discussed above, novel OTO-transport parameters can be measured by considering OTOCs involving commutators, in particular $g_0 - g_2$ defined in Eq.~\eqref{eq_4pt_basis}. This observable is shown in Fig.~\ref{fig_num_lambda12}: we first observe that $g_0 - g_2 \ll g_0$ at late time, confirming the heuristic argument that operators approximately commute at late times (which is substantiated by the EFT). Furthermore, $\Re(g_0 - g_2)$ indeed appears to have a rapid power-law decay at lates times consistent with the EFT prediction, $1/t^{7/2}$. The coefficient of this decay is a measurement of the OTO-transport parameters $\lambda_{1,2}$. There also appears to be an intermediate $1/t^{3/2}$ behavior in this observable (see also Fig.~\ref{Fig-Ising-finiteT}(c)) which we do not have an explanation for---it would be interesting to understand if this intermediate regime is generic. 

Finally, we perform a thorough test of the EFT predictions by considering all of the 6 predictions for power-law behavior given in Table~\ref{tab:scaling_correlators}. We find good agreement between EFT and numerics for all these power laws (with the exception of the intermediate regime in $\Re (g_0 - g_2)$ mentioned above), see Fig.~\ref{Fig-Ising-finiteT}. Note that all log-log plots show the absolute value of the corresponding correlator. While hydrodynamic tails with large exponents are often difficult to match between numerics and theory \cite{Glorioso:2020loc,matthies2025thermalizations}, it is encouraging that the $1/t^{3/2}$ and $1/t^{5/2}$ power laws show good agreement with theory. App.~\ref{app_num} shows similar results for different parameters and temperatures.

\section{Discussion}

Our effective field theory revealed the existence of novel transport parameters in diffusive quantum many-body systems that are only visible in OTOCs. These parameters were measured numerically in several one-dimensional models in Sec.~\ref{sec_numerics}. It should be possible to evaluate these OTO-transport parameters analytically in tractable models, such as holographic systems or Brownian models \cite{Ogunnaike:2023qyh}. Recent experimental breakthroughs have also made OTOCs accessible in experimental platforms including trapped ions \cite{Garttner:2016mqj,Landsman:2018jpm,Joshi:2020quh}, NMR on large molecules \cite{PhysRevX.7.031011, Zhang:2025qph}, and superconducting quantum circuits \cite{Braumuller:2021cic,Abanin:2025rbz}, where our predictions could be tested. 

It would be interesting to understand the interplay of diffusive dynamics in OTOCs with information dynamics and the butterfly front, see Refs.~\cite{Khemani:2017nda,Agarwal:2022teo} for observations in this direction. The butterfly front is often discussed through the lens of operator weights, which satisfy a conservation law and therefore have a hydrodynamic-like description \cite{Khemani:2017nda,Nahum2018operator}. However it is unclear if this description applies to OTOCs themselves at finite temperature. As discussed in Sec.~\ref{ssec_regime}, it is challenging at first glance to maintain control in the EFT in the regime where one expects a butterfly front. Furthermore, while butterfly fronts are very sensitive to the presence of dephasing noise \cite{Jacoby:2024ftp}, we expect our OTO-transport parameters to be more robust. \new{It would also be interesting to understand how to unify our EFT with other EFTs proposed in the context of large-$N$ holographic QFTs which capture Lyapunov growth \cite{Blake:2017ris, Choi:2023mab,Gao:2023wun}. See Refs.~\cite{Grozdanov:2017ajz, Choi:2020tdj,Asplund:2025nkw} for other connections between hydrodynamics and the butterfly effect in large-$N$ holographic QFTs.}

It would also be interesting to identify classical versions of the OTOC---other than the obvious Poisson bracket which is dominated by the butterfly effect and Lyapunov growth---where analogs of our OTO-transport parameters may appear \cite{PhysRevLett.121.024101,Bilitewski:2020xkb,Kumar_2020}. See also Refs.~\cite{Gopalakrishnan:2018wqv,Suchsland:2022mct} for further rich phenomenology of OTOCs in different contexts.


On a formal note, it is somewhat surprising that OTOCs can reveal transport parameters that are invisible in TOCs, given that they can both be obtained by analytic continuations of the Euclidean (imaginary time) Green's function. Of course, hydrodynamic transport parameters are only defined in an asymptotic late time expansion, so that this is not a contradiction.

It is also interesting to note that, unlike equilibrium EFTs, EFTs for real time thermal dynamics are `observable-dependent'. For example, the appropriate EFT for a given system will feature more and more switchbacks (Fig.~\ref{fig_contours}) if one is interested in higher order OTOCs. We have seen that more complex observables are sensitive to additional transport parameters. It would be interesting to understand whether more sophisticated observables could feature entirely different universality classes. Refs.~\cite{Winer:2020gdp,Gopalakrishnan:2025lfx} hint that this may indeed occur; for the observables considered there, the appropriate EFT includes a $\phi_r^2$ term that would otherwise be forbidden by a latest-time condition in the conventional 1-CTP EFT.

There are many possible extensions to our EFT, including to systems with sound modes, non-abelian symmetries, anomalous symmetries, higher-form or other generalized symmetries. It would be particularly interesting to study the effect of large hydrodynamic fluctuations on OTOCs, as arise for example in the KPZ universality class that accompanies sound modes in one spatial dimension.

\section*{Data availability}
The data that support the findings of
this article are openly available \cite{data}.

\section*{Acknowledgements}

We thank Clay C\'ordova, Richard Davison, Eren Firat, Sarang Gopalakrishnan, Yingfei Gu, David Huse, Jong-Yeon Lee, Mark Mezei, Marvin Qi, Mukund Rangamani, Ramanjit Sohal, and Dam Thanh Son for insightful discussions. RM and LD are supported by an NSF award No.~PHY2412710 and a Sloan Fellowship. JW acknowledges support from DFG, under Grant No. 531128043, as well as under Grant
No.\ 397107022, No.\ 397067869, and No.\ 397082825 within the DFG Research
Unit FOR 2692, under Grant No.\ 355031190.
Additionally, we greatly acknowledge computing time on the HPC3 at the University of Osnabr\"{u}ck, granted by the DFG, under Grant No. 456666331.
SP acknowledges
support by the Deutsche Forschungsgemeinschaft
(DFG, German Research Foundation) under Germany’s Excellence Strategy - Cluster of Excellence Matter and Light for Quantum Computing
(ML4Q) EXC 2004/1 -390534769, and DFG Collaborative Research Center (CRC) 183 Project
No. 277101999 - project B02.

\bigskip

\section*{Appendix}
\appendix

\makeatletter
\renewcommand\subsection{\@startsection{subsection}{2}{0pt}%
  {1.0ex plus .5ex minus .2ex}%
  {0.5ex plus .2ex}%
  {\normalfont\normalsize\itshape\centering\medskip}}
\makeatother

\section{Review of the Schwinger-Keldysh EFT approach to fluctuating hydrodynamics}\label{app_review_Z2}

Fluctuating hydrodynamics is a time-honored framework that is expected to describe the late time dynamics of any interacting many-body system at nonzero temperature \cite{Martin:1973zz, dominicis1976techniques, Janssen:1976qag}. Compared to the `classical' hydrodynamics of differential equations satisfied by conserved densities (the Navier-Stokes equation, the diffusion equation, etc.), it recognizes the need to supplement the description with additional dynamical `noise' fields that capture fluctuations in the thermal state. Its early successes include the description of hydrodynamic long-time tails \cite{Alder:1970zza,PhysRevLett.25.1254}, and the discovery of novel dissipative universality classes beyond diffusion \cite{PhysRevA.16.732}. More recently, the reformulation of fluctuating hydrodynamics in terms of an EFT on a Schwinger-Keldysh contour \cite{Crossley:2015evo,Haehl:2015foa,Jensen:2018hse}, and the subsequent realization that the long-lived degrees of freedom could be interpreted as the Nambu-Goldstone modes of SWSSB \cite{Ogunnaike:2023qyh,Akyuz:2023lsm}, have allowed to recast fluctuating hydrodynamics in the more familiar language of EFTs for symmetry broken phases in equilibrium. While this modern reformulation is equivalent to the previous approaches %
	\footnote{One exception is in `superplanckian' theories of hydrodynamics, with EFT cutoff is parametrically larger than temperature $1/\tau_{\rm eq} \gg T/\hbar$; in this situations, the modern approach allows to impose KMS exactly with specific nonlocal terms in the action. However, generic quantum many-body systems are expected to satisfy $\tau_{\rm eq}\gtrsim \hbar/T$ \cite{Delacretaz:2023pxm}}, 
its versatility allows for generalizations beyond conventional observables, such as the ones pursued in this paper. In this section, we briefly review this modern approach to fluctuating hydrodynamics (see also \cite{Liu:2018kfw} for a review).

Consider the generating functional $Z[A^1,A^2]$ defined in \eqref{eq_Z2}. It satisfies several conditions:
\begin{enumerate}
\item
	$Z[A,A] = 1$ \quad (collapse)
\item
	$Z[A^1,A^2]^* = Z[A^2,A^1]$ \quad (unitarity)
\item
	$Z[A^1,A^2] = Z[A^1(-t+i\beta),A^2(-t)]$ \quad (KMS)
\end{enumerate}
The notation $Z[A^1(-t+i\beta),\ldots]$ means that the first argument is a function $\tilde A^1$ defined by $\tilde A^1(t) = A^1(-t+i\beta)$.
KMS involves dropping a contour of length $i\beta$ at $t=\pm \infty$, and acting with time-reversal. As discussed below Eq.~\eqref{eq_KMS}, time-reversal symmetry is not necessary to impose KMS, but we assume it for simplicity. In the Keldysh basis, these conditions read 
%
\begin{enumerate}
\item
	$Z[A^r,A^a=0] = 1$ \quad (collapse)
\item
	$Z[A^r,A^a]^* = Z[A^r,-A^a]$ \quad (unitarity)
\item
	$Z[A^r,A^a] = Z[\tilde A^r,\tilde A^a]$ \quad (KMS)
\end{enumerate}
with
\begin{align}
\tilde A^r(-t) 
	&= \cosh \frac{i\beta \partial_t}{2} A^r(t) + \frac{1}{2} \sinh \frac{i\beta \partial_t}{2} A^a(t)\, , \\ 
\tilde A^a(-t) 
	&= \cosh \frac{i\beta \partial_t}{2} A^a(t) + 2 \sinh \frac{i\beta \partial_t}{2} A^r(t)\, .
\end{align}
This expression is simpler in frequency space:
\begin{align}
\tilde A^I(\omega) &= h^I{}_J(\omega) A^J(-\omega)\, ,  \\
h^I{}_J(\omega) &=
\left(\begin{smallmatrix}\cosh \frac{\beta\omega}{2}&-\frac12 \sinh \frac{\beta\omega}{2}\\
-2 \sinh \frac{\beta\omega}{2}&\cosh \frac{\beta\omega}{2}\end{smallmatrix}\right)\, .
\end{align}
Note that KMS$^2 = h(-\omega)h(\omega) = \mathds 1$.

Condition 1 (collapse) states that if the sources on the two legs are identified, the legs collapse and produce $\Tr \rho = 1$. There is a stronger version of this statement: if the two sources are only identified after a certain time $t_o$, the dependence on that source drops out. In the Keldysh basis, if $A^a$ is zero beyond a time $t_o$, then the dependence on $A^r(t>t_o)$ drops:
\begin{equation}\label{eq_Zlatesttime}
Z[A^r,A^a\Theta (-t+t_o)] = Z[A^r\Theta(-t+t_o),A^a\Theta (-t+t_o)]\, .
\end{equation}
We refer to this stronger version of the collapse constraint as the `latest time condition', since it implies that correlators involving a $\phi_a$ field at the latest time must vanish.

SWSSB implies that one expects this generating functional to be described by a local EFT at late times
\begin{equation}
Z[A]
	= \int D\mu^r D \phi^a\, e^{iS[\mu^r,\phi^a,A]}\, .
\end{equation}
One can translate the conditions on $Z$ into constraints on the effective action (see App.~\ref{app_Z_to_S}):
\begin{enumerate}
\item
	$S[\phi^r,\phi^a=0]=0$ \quad (collapse)
\item
	$S^*[\phi^r,\phi^a] = - S[\phi^r,-\phi^a]$ \quad (unitarity)
\item
	$S[\phi^r,\phi^a]
	= S[\tilde \phi^r,\tilde \phi^a]$ \quad (KMS)
\item
	$\Im S\geq 0$ \quad (stability)
\end{enumerate}
These conditions are simpler to express in terms of a Keldysh partner $\phi^r$ to the SWSSB Nambu-Goldstone field $\phi^a$. However, we should keep in mind that the true degree of freedom entering the EFT is $\mu^r = \dot \phi^r$. 

Let us now construct the action $S$ satisfying these constraints, working perturbatively in derivatives and fields. At the Gaussian level
\begin{equation}
S_{(2)} = \frac12 \int_p \phi^i_{-p} M_{ij}(p) \phi^j(p)\, , 
\end{equation}
collapse imposes $M_{rr}=0$, and unitarity imposes $M_{aa}(p) \in \mathbb R$ and $M_{ra}(p)^*=M_{ra}(-p)$. The KMS condition $M(p) = h(\omega)^T M^T(p) h(-\omega)$ is solved by
\begin{equation}
M_{aa}(p) = \Im M_{ra}(p) \coth \frac{\beta \omega}{2}\, .
\end{equation}
Expanding now in derivatives, and recognizing that SWSSB implies that only $\mu^r = \dot \phi^r$ should enter the action:
\begin{equation}
M_{ra}(p)= \chi (- i\omega) (i\omega - D k^2 + \cdots), 
\end{equation}
which leads to the action 
\begin{equation}\label{eq_app_S2_1CTP}
S_{(2)} = \chi \int_{tx} \dot \phi^a\mu^r +  D \nabla\phi^a(i T \nabla\phi^a - \nabla\mu^r)  + \cdots .
\end{equation}
\new{The leading order quadratic action has two independent coefficients $\chi$ and $D$ (or alternatively $\chi$ and $\sigma=\chi D$).}

We now turn to the cubic action:
\begin{equation}
S_{(3)}
	= \frac1{3!} \int_{p_1p_2p_3} \!\!\delta_{\Sigma_i p_i} M_{ijk}(p_1,p_2,p_3) \phi^i_{p_1}\phi^j_{p_2}\phi^k_{p_3}\, ,
\end{equation}
with $\delta_{\Sigma_i p_i}\equiv (2\pi)^{d+1} \delta^{d+1}(\sum_i p_i)$.
The collapse condition implies that $M_{rrr}=0$, leaving $M_{rra},M_{raa},M_{aaa}$ as independent components, and unitarity gives the reality condition
%
\begin{equation}
\begin{split}
M_{rra}^*( p )&=M_{rra}( -p )\, , \quad
M_{raa}^*( p )=-M_{raa}( -p )\, , \\
M_{aaa}^*( p )&=M_{aaa}( -p )\, .
\end{split}
\end{equation}
Finally, KMS requires
\begin{equation}
M_{ijk}(\{-p\})
	= - M_{i'j'k'}(\{p\}) h^{i'}{}_i(p_1) h^{j'}{}_j(p_2) h^{k'}{}_k(p_3)\, .
\end{equation}
These constraints can be solved for $M_{aaa}$ and $M_{raa}$ in terms of $M_{rra}$. For example one has
\begin{equation}\label{eq_m3_KMS}
\begin{split}
&M_{raa}(p_1,p_2,p_3)
	= \frac{-\frac12\sinh \frac{\beta \omega_1}{2}}{\sinh \frac{\beta \omega_2}{2}\sinh \frac{\beta \omega_3}{2}} M_{rra}(-p_2,-p_3,-p_1)\\
	&+ \frac{\coth \frac{\beta \omega_2}{2}}{2}M_{rra}(p_1,p_2,p_3)
	+ \frac{\coth \frac{\beta \omega_3}{2}}{2}M_{rra}(p_1,p_3,p_2)\, , 
\end{split}
\end{equation}
and a similar but more complicated expression for $M_{aaa}$. Let us now expand in derivatives, keeping in mind that the SWSSB pattern requires $\phi_r$ to enter with time derivatives. To leading order in the diffusive scaling, we expect
\begin{equation}
M_{rra}(p_1,p_2,p_3) = \omega_1 \omega_2 (c_1 i\omega_3 + c_2 k_3^2) + \cdots \, .
\end{equation}
Some other candidates are equivalent after integration by parts. Unitarity implies that $c_1,c_2\in \mathbb R$. They will have no sign constraint as they are not the leading dissipative terms. Eq.~\eqref{eq_m3_KMS} then gives
\begin{equation}
\begin{split}
\beta M_{raa}(p_1,p_2,p_3)
	&\simeq -2c_2 \omega_1 k_2\cdot k_3\, .
\end{split}
\end{equation}
Returning to real space, we find the leading order cubic action
\begin{equation}
S_{(3)}
	= \int_{tx}\,  c_1 \mu_r^2 \dot \phi_a + c_2 \mu_r\nabla\phi_a\left( iT  \nabla\phi_a - \nabla\mu_r\right)\, .
\end{equation}
As discussed in \cite{Chen-Lin:2018kfl,Jain:2020zhu,Delacretaz:2023ypv}, the parameters $c_1,\,c_2$ entering the cubic action are equal to derivatives of the linear response parameters with respect to chemical potential: $2c_1 = \chi'\equiv d\chi(\mu)/d\mu$ and $c_2 = \sigma' \equiv d\sigma(\mu)/d\mu$. One can show more generally that higher order terms in the action follow this pattern, and the nonlinear EFT to leading order in gradients takes the form
\begin{equation}
S
	 = \int_{tx} \dot \phi^an(\mu^r) +  \sigma(\mu^r) \nabla\phi^a(i T \nabla\phi^a - \nabla\mu^r)  + \cdots .
\end{equation}
It is an interesting prediction of the EFT of diffusion that nonlinear response in {\em time-ordered} correlators at late times is entirely fixed by data available from linear response at various densities, i.e.~$\chi(\mu)$ and $\sigma(\mu)$. For example, three-point functions are given by a universal functional form at late times that only depends on $\chi, \sigma$ and their first derivatives $\chi',\sigma'$, see \cite{Delacretaz:2023ypv}. Similarly, four-point functions that can be captured on a 1-CTP contours will depend on the additional parameters $\chi'',\sigma''$.

\section{From generating functional to effective action}\label{app_Z_to_S}

The definition of the generating functional \eqref{eq_Z4} immediately implies several simple properties that it must satisfy, which we refer to as collapse \eqref{eq_collapse}, unitarity \eqref{eq_unitarity}, and KMS \eqref{eq_KMS} constraints. In this appendix, we explain how to turn these conditions into constraints on the effective action.

\subsection{KMS and unitarity constraints}\label{sapp_KMS}
\medskip

Consider the path integral respresentation of the generating functional
\begin{equation}\label{eq_app_ZfromS}
Z[A] = \int D\phi_{1,2,3,4} \, e^{i S[\phi] + i \int A_I g^{IJ}\phi_I}\, , 
\end{equation}
where $g^{IJ} = {\rm diag} (1,-1,1,-1)$ is a Minkowski-like `metric' in field space. One appeal of introducing this metric is that a 2-CTP action of the form $S\supset \sum_I (-1)^{I} S_0(\phi_I) = \int \dot \phi_I g^{IJ}\dot \phi_J$ is certainly allowed, so that symmetries of the EFT that act on the legs will form a subgroup of $SO(2,2)$. KMS is such a symmetry: from \eqref{eq_KMS} we see that it acts on the sources as $A_I \to \tilde A_I$ with
\begin{subequations}\label{eq_2CTP_KMS}
\begin{align}
	(\tilde A_{K^2})_I(\omega)&= (h_{K^2})_I{}^J(\omega) A_J(\omega)\, ,\\
	(\tilde A_{TK})_I(\omega)&= -(h_{TK})_I{}^J(\omega) A_J(-\omega)\, ,
\end{align}
\end{subequations}
with, in the 1234 basis,
\begin{align}
h_{K^2}
	&= 
\left(\begin{smallmatrix}
	&&1\\
	&&&1\\
	e^{\beta\omega}\\
	&e^{\beta\omega}
\end{smallmatrix}\right),&
h_{TK}
	&= \left(\begin{smallmatrix}
	e^{-\beta\omega}\\
	&&&1\\
	&&1\\
	&1
\end{smallmatrix}\right).
\end{align}
Eq.~\eqref{eq_app_ZfromS} shows that $Z$ is the fourier transform of $e^{iS}$. The symmetry of $Z$ is thus equivalent to the symmetry of $S$, which gives Eq.~\eqref{eq_action_KMS} from the main text.

The unitarity condition can be treated similarly. Eqs.~\eqref{eq_unitarity} and \eqref{eq_app_ZfromS} lead to 
\begin{equation}
S^*[\phi_1,\phi_2,\phi_3,\phi_4] = -S[\phi_4,\phi_3,\phi_2,\phi_1]\, , 
\end{equation}
which in the Keldysh basis \eqref{eq_KeldyshRotation_Z4} gives \eqref{eq_action_unitarity}.

\subsection{Collapse constraint at tree-level}\label{sapp_collapse}
\medskip
\new{Next, we want to show how the collapse constraints \eqref{eq_collapse} on the generating functional lead to constraints on the effective action \eqref{eq_action_collapse}. Our argument is an $n$-CTP generalization of the argument presented in \cite{Crossley:2015evo}. So, the only difference is the presence of additional sources $A_k$ and fields $\phi^k$ in higher CTP. }

\new{We first note that we can rewrite the constraints \eqref{eq_collapse} in a new basis such that setting two sources equal is the same as setting a source to be zero in the new basis. In other words, if we consider the constraint $A_1=A_2=A$ as an example, we can rewrite that in the new basis as $\tilde{A}_1\equiv A_1-A_2=0$ and $\tilde{A}_2\equiv\frac{A_1+A_2}{2}=A$. Then the constraint \eqref{eq_collapse} tells us that when $\tilde{A}_1=0$, the generating functional no longer depends on $\tilde{A}_2$. More generally, upon a linear transformation of the sources, the collapse constraints \eqref{eq_collapse} can be written as:
\begin{equation}\label{eq_Zcollapse}
    Z_{2n}[A_i=0, A_j, A_k,\cdots]=Z_{2(n-1)}[A_k,\cdots]
\end{equation}
At tree level, this leads to a corresponding constraint on the EFT
(note the raised indices, $\phi^k\equiv g^{kl}\phi_l$):
\begin{equation}\label{eq_Scollapse}
	S_{2n}[\phi^i,\phi^j = 0, \phi^k,\ldots] = S_{2n-1}[\phi^k,\ldots]\, .
\end{equation}
Our goal is to now show that this follows from \eqref{eq_Zcollapse}.   }

Let us show that the LHS of \eqref{eq_Scollapse} is independent of $\phi^i$. At tree-level, the generating function is related to the on-shell action:
\begin{equation}\label{eq_Zonshell}
\begin{split}
Z[A_i,A_j,\ldots]
	&= \int D\phi_i\cdots e^{i \left(S[\phi] + A_i \phi^i\right)}\\
	&\approx e^{i \left(S[\bar \phi] + A_i \bar\phi^i\right)}\, , 
\end{split}
\end{equation}
where $\bar\phi(A)$ is the solution to the equation of motion
\begin{equation}
0 = \frac{\delta S[\phi]}{\delta \phi^i} + A_i\, .
\end{equation}
Now \eqref{eq_Zcollapse} implies in particular
\begin{equation}
\left.\frac{\delta \new{Z_{2n}}[A]}{\delta A_j}\right|_{A_i=0}
	=0\, . 
\end{equation}
Using \eqref{eq_Zonshell} this implies
\begin{equation}\label{eq_phib_id}
\begin{split}
0 &= \left[\left(\left.\frac{\delta S[\phi]}{\delta\phi_i}\right|_{\bar \phi} + A_i\right) \frac{\delta \bar\phi_i[A]}{\delta A_j} + \bar\phi^j\right]_{A_i=0}\\
	&=  \bar\phi^j[A_i=0,A_j,A_c,\ldots]\, .
\end{split}
\end{equation}
where in the last step we used the fact that $\bar\phi$ solves the equations of motion. Now consider the inverse function of $\bar\phi(A)$, namely
\begin{equation}
\bar A_i[\phi]
	\equiv -\frac{\delta S[\phi]}{\delta \phi^i}\, .
\end{equation}
\new{Since, $\bar{A}_i[\bar{\phi}[A]]=A_i$ by definition, } Eq.~\eqref{eq_phib_id} implies $\bar A_i[\phi^i,\phi^j=0,\phi^k,\ldots]=0$, which shows that $S[\phi^i,\phi^j=0,\phi^k,\ldots]$ is independent of $\phi^i$ as claimed. Note that the map between $Z$ and $S$ is assumed to be injective. This finally leads to the collapse condition on the effective action quoted in the main text, Eq.~\eqref{eq_action_collapse}.

\subsection{All-order generalization and latest time condition}\label{sapp_collapse_loop}
\medskip
\new{The argument in the previous section followed from a saddle-point approximation \eqref{eq_Zonshell} and thus holds at tree-level. The leading order OTOCs at late times are obtained from tree-level diagrams in the EFT; the collapse constraint on the action obtained in App.~\ref{sapp_collapse} in the tree-level approximation is thus sufficient. Nevertheless, it can be shown that this condition is also sufficient to all order in loops. In other words, imposing collapse on the EFT \eqref{eq_action_collapse} is sufficient to capture collapse of the generating functional \eqref{eq_collapse} even beyond tree level. This was shown for the 1-CTP case in Ref.~\cite{Gao:2018bxz}, which also showed that the `latest time condition,' described below, follows from imposing collapse condition on the EFT. A key ingredient was the retarded nature of the $G_{ar}$ propagator.  }

\new{
The collapse constraint involves collapsing the field for all times. For example, in the 1-CTP case, $Z[A_a=0, A_r]=1$ where $A_a=0$ for all times. A consequence of this is that $\langle\phi^a\phi^a\cdots\phi^a\rangle=0$. The latest time condition \eqref{eq_Zlatesttime} on the other hand involves setting $A_a(t)=0$ after time $t=t_{o}$,  such that the correlator with $\phi^a(t)$ field placed at the latest time is zero i.e.
\begin{equation}\label{eq_1ctp-latesttime}
    \langle\phi^a(t_{m+1})\phi^{\alpha_m}(t_m)\cdots\phi^{\alpha_1}(t_1)\rangle=0 \quad t_{m+1}>t_i
\end{equation}
where $\phi^{\alpha_i}$ can be either $(\phi^a,\phi^r)$. On a general $n$-CTP contour, there is more than one latest time condition---one for each pair of legs. In addition, there are also {\em earliest} time conditions. 
Denoting $\phi^{a_i}\equiv \phi_i-\phi_{i+1}$, $\phi^{i_\alpha}$ as fields also on legs $(i,i+1)$ and $\phi^{k_\alpha}$ as fields on all other legs, the latest time condition can be written as 
 \begin{equation}\label{eq_latesttime-i-even}
     \langle \phi^{a_i}(t)\phi^{i_1}(t_1)\cdots\phi^{i_l}(t_l)\phi^{k_1}\cdots \phi^{k_m}\rangle=0\quad  \forall\, t>t_i 
\end{equation}
 when $i\in\{1,3,\cdots 2n-1\}$ in $\phi^{a_i}\equiv \phi_i-\phi_{i+1}$. The earliest time condition holds when the $\phi^{a_i}\equiv \phi_i-\phi_{i+1}$ fields lie on legs $i\in\{2,4,\cdots 2n-2\}$,
 \begin{equation}\label{eq_latesttime-i-odd}
  \langle \phi^{a_i}(t)\phi^{i_1}(t_1)\cdots\phi^{i_l}(t_l)\phi^{k_1}\cdots \phi^{k_m}\rangle=0\quad  \forall\, t<t_i 
\end{equation}
Note that the equations are true for any time ordering of the spectator $\phi^{k_\alpha}$ fields. The proof that \eqref{eq_latesttime-i-even} and \eqref{eq_latesttime-i-odd} hold at all loop orders by imposing collapse conditions \eqref{eq_Scollapse} on the EFT follows from a simple generalization of the proof in \cite{Gao:2018bxz}.}

\medskip

\subsection{A comment on positivity conditions}\label{sapp_positivity}
\medskip

When hydrodynamic correlators are generated by a path integral as in this paper, it is natural to require $\Im S\geq 0$ to guarantee its convergence \cite{Crossley:2015evo,Glorioso:2016gsa,Jensen:2018hse,Liu:2018kfw}. However, we will argue here that this is too conservative a requirement, and leads to incorrect bounds on transport parameters. Let us illustrate this with the usual 1-CTP diffusion EFT \eqref{eq_app_S2_1CTP}, considering one of its higher derivative corrections
\begin{align}
S = \chi \int_{tx} \nabla_0\phi^a\mu^r &+  D \nabla_i\phi^a(i T \nabla_i\phi^a - \partial_i\mu^r  \new{+} E_{i,r}) \notag \\
	&+ \alpha \nabla_0 \phi^a (iT \nabla_0 \phi^a - \partial_0 \mu^r)\, ,
\end{align}
where we have coupled to background gauge fields and let $\nabla_\mu \phi^a \equiv \partial_\mu \phi^a + A_\mu^a$, with $\partial_0 \equiv \partial_t$. On one hand, requiring $\Im S\geq 0$ for any field configuration in the path integral would require $\alpha\geq 0$ (choose $\phi^a = \phi^a(t)$). On the other hand, the extra term with coefficient $\alpha$ is irrelevant in the diffusive scaling $\omega \sim k^2$; we therefore do not expect it to satisfy a bound from positivity of dissipation since a positive conductivity $\chi D>0$ will dominate spectral densities. For example, the retarded Green's function for charge density from this action is
\begin{equation}
G^R_{nn}(\omega,k)
	= \frac{\chi D k^2 (1 + i\omega \alpha)}{-i\omega + D k^2 + \alpha \omega^2}\, ,
\end{equation}
so that $\omega\Im G^R\geq 0$ requires $\chi D\geq 0$ but sets no constraint on $\alpha$ within the regime of validity of hydrodynamics. Higher derivative corrections to diffusion including $\alpha$ were measured in a Floquet system \cite{Michailidis:2023mkd}; it would be interesting to see whether $\alpha < 0$ can be realized in models.

Another counterexample is provided by the 2-CTP EFT, where the cubic action has an imaginary part (see Eq.~\eqref{eq_app_sigmap})
\begin{align}
\Im \mathcal L_{(3)}
	= \frac{T\sigma'}{4} 
	&\bigl[\nabla\phi_A \left(\nabla\phi_A \dot \phi_R + \nabla\phi_+\dot \phi_- + \nabla\phi_- \dot \phi_+\right)\notag\\
	&
	-\nabla\phi_- \nabla\phi_+ \dot \phi_A\bigr]\, .
\end{align}
Setting $\phi_A = \phi_A(t)$, the last term remains and is non-sign definite. Imposing $\Im S\geq 0$ would therefore lead to the incorrect conclusion that $\sigma'=0$.


\section{Details of the 2-CTP EFT}
\subsection{OTOC parameters}\label{app_terms}

Starting with the most general quartic action, 
\begin{equation}
    S_{(4)}
	= \frac{1}{4!} \int_{p_1p_2p_3p_4} \!\delta_{\Sigma_i p_i} M_{IJKL}(\{p\}) \phi_{p_1}^I \phi_{p_2}^J \phi_{p_3}^K\phi_{p_4}^L\, , 
\end{equation}
we impose collapse, KMS and unitarity constraints. While collapse and unitarity are easy to impose, solving the equations for the full KMS transformations is hard compared to the quadratic case. Instead, we impose KMS perturbatively by expanding \eqref{eq_2CTP_KMS} in $\beta\omega$. In other words, we impose that the action be invariant under the following two transformations of the fields:
First is the expanded version of $K^2$,
\begin{equation}
\begin{split}
\mu_R &\to \mu_R + \frac{i\beta}{2}\partial_t \nabla_0 \phi_+\\
\nabla_\mu \phi_A &\to \nabla_\mu \phi_A + \frac{i\beta}{2}\partial_t \nabla_\mu \phi_-\\
\nabla_\mu \phi_+ &\to -\nabla_\mu \phi_+ - \frac{i\beta}{2}\left(\partial_\mu \mu_R + F_{0\mu,R}\right)\\
\nabla_\mu \phi_- &\to -\nabla_\mu \phi_- - \frac{i\beta}{2}\partial_t \nabla_\mu \phi_A
\end{split}
\end{equation}
\new{where $F_{\mu0,R}=\delta_\mu^iE_{i,R}$ is the electric field}. And, the second one is $TK$,
\begin{equation}
\begin{split}
\mu_R(t)  &\to \mu_R(-t)  + \delta_0\phi(-t)\\
\nabla_\mu \phi_I &\to \eta_{\mu}(- \nabla_\mu \phi_I(-t) - \delta_\mu\phi(-t))
\end{split}
\end{equation}
with
\begin{equation}
\delta_\mu \phi = \frac{i\beta}{4} \left[\partial_t \nabla_\mu (\phi_A + \phi_+ + \phi_-) + \partial_\mu \mu_R + F_{0\mu,R}\right] \, .
\end{equation}
and $(\eta_0,\eta_i)=(-1,1)$.
On imposing these constraints,  at leading order in scaling, we find three terms- one of which collapses to the 1-CTP $\sigma''$ term. The other two terms collapse to zero (at the order we work at), meaning that these terms are potential new otoc transport parameters. We call their coefficients $\lambda_{1,2}$. They have the following form at leading order in scaling,  
\begin{equation}\label{eq_app_l1}
    \begin{split}
     \frac{-i\lambda_1}{\beta}\Bigg(   &\dot{\phi}_-^2 \left(\partial{\phi}_A^2-\partial{\phi}_+^2\right)+\partial{\phi}_-^2 \left(\dot{\phi}_A^2-\dot{\phi}_+^2\right)\\
     &+\frac{i\beta}{2}\Big(-\dot{\phi}_- \partial{\phi}_- (\ddot{\phi}_A \partial{\phi}_-+2 \dot{\phi}_A \partial{\dot{\phi}}_-)\\
     &+\partial{\phi}_-^2 (\dot{\phi}_A \ddot{\phi}_R-\dot{\phi}(p) (\ddot{\phi}_-+\ddot{\phi}_R))\\
     &+\dot{\phi}_-^2 (\partial{\phi}_A-\partial{\phi}_+) (\partial{\dot{\phi}}_-+\partial{\mu_R}\Big) \Bigg)
    \end{split}
\end{equation}
And, 
\begin{equation}\label{eq_app_l2}
    \begin{split}
    \frac{-i\lambda_2}{\beta}\Bigg(&\Big(\dot{\phi}_- \partial{\phi}_- (\dot{\phi}_A \partial{\phi}_A-\dot{\phi}_+ \partial{\phi}_+)\Big)\\
    &+\frac{i\beta}{8}\Big(\dot{\phi}_- (2 \partial\phi_- (\ddot{\phi}_R (\partial\phi_A-\partial\phi_+)\\
    &+(\dot{\phi}_A-\dot{\phi}_+) (\partial\dot{\phi}_-+\partial\mu_R))\\
    &+\dot{\phi}_- (\partial\phi_- (\partial\dot{\phi}_+-\partial\dot{\phi}_A)+\partial\dot{\phi}_- (\partial\phi_+-\partial\phi_A)))\Big)\Bigg)
 \end{split}
\end{equation}
We also write here the leading order expressions for the terms that collapse to regular 1-CTP terms. First, we have the cubic terms that collapse to the $\chi'$ and $\sigma'$ terms respectively:
\begin{equation}
    \begin{split}
        \frac{\chi\chi'}{32}\Big(&\dot{\phi}_A \left(\dot{\phi}_-^2+\mu_R^2\right)+2 \dot{\phi}_- \dot{\phi}_+ \mu_R\Big)
    \end{split}
\end{equation}
\begin{equation}\label{eq_app_sigmap}
    \begin{split}
      \frac{i\chi\sigma'}{4\beta}\Big(&\dot{\phi}_+\partial{\phi_A}  \partial{\phi_-}+\partial{\phi_+} (\partial{\phi_A} \dot{\phi}_--\dot{\phi}_A \partial{\phi_-})+(\partial{\phi_A})^2 \mu_R\\
        &-\frac{i\beta}{4}\big((\dot{\phi}_A-\dot{\phi}_+)\partial{\phi_-}  (\partial\dot{\phi}_-+\partial\mu_R)\\
        &+\mu_R (\partial{\phi_-} (\partial\dot{\phi}_A-\partial\dot{\phi}_+)-\partial{\phi_A} (\partial\dot{\phi}_-+\partial\mu_R))\\
        &+\dot{\phi}_- (\partial{\phi_-} (\partial\dot{\phi}_A-\partial\dot{\phi}_+)\\
        &+\partial\dot{\phi}_- (\partial{\phi_A}-2 \partial{\phi_+})-\partial{\phi_A}\partial\mu_R))\big)\Big)
    \end{split}
\end{equation}
Similarly, the quartic terms that collapse to $\chi''$ and $\sigma''$ have the following form at leading order:
\begin{equation}
    \begin{split}
        \frac{\chi^2\chi''}{384}\Big( 3 \dot{\phi}_A \dot{\phi}_-^2 \mu_R+\dot{\phi}_A \mu_R^3+3 \dot{\phi}_- \dot{\phi}_+ \mu_R^2+\dot{\phi}_-^3 \dot{\phi}_+ \Big)
    \end{split}
\end{equation}
\begin{equation}
    \begin{split}
        \frac{i\chi^2\sigma''}{32}&\Big(\partial{\phi_A} \left(\partial{\phi_A} \left(\dot{\phi}_-^2+\mu_R^2\right)+2 \dot{\phi}_+ \mu_R \partial{\phi_-}\right)\\
        &+2 \mu_R \partial{\phi_+} (\partial{\phi_A} \dot{\phi}_--\dot{\phi}_A \partial{\phi_-})\\
        &+\frac{i\beta}{4}\Big(2 \dot{\phi}_- \big(\dot{\mu}_R \partial{\phi_-} (\partial{\phi_+}-\partial{\phi_A})\\
        &+\mu_R (\partial{\phi_-} (\partial\dot{\phi}_+-\partial\dot{\phi}_A)\\
        &+\partial{\phi_A} (\partial\mu_R-\partial\dot{\phi}_-)+2 \partial\dot{\phi}_- \partial{\phi_+})\big)\\
        &+\dot{\phi}_-^2 (\partial{\phi_A} (2 \partial\dot{\phi}_-+\partial\mu_R)-\partial\dot{\phi}_- \partial{\phi_+})\\
        &+\mu_R (\mu_R (\partial{\phi_-} (\partial\dot{\phi}_+-\partial\dot{\phi}_A)+\partial{\phi_A} (\partial\dot{\phi}_-+\partial\mu_R))\\
        &-2 \partial{\phi_-} (\dot{\phi}_A-\dot{\phi}_+) (\partial\dot{\phi}_-+\partial\mu_R))\Big)\Big)
    \end{split}
\end{equation}

\subsection{Coupling to background sources}
Note that the procedure above gives us an action without any background sources. To calculate correlators, we will need the fully gauge-invariant action with background sources. To introduce sources, we first note that from the symmetry breaking pattern, the relevant degrees of freedom are the three Nambu-Goldstone modes $(\phi_A,\phi_+,\phi_-)$ and a density field $\mu_R$.
We can couple background sources to the Goldstones in a gauge-invariant way as follows: $\nabla_\mu\phi^I=\partial_\mu\phi^I+A_\mu^I$. For the density, we note that gauge invariance and KMS require that $\partial_i\mu_R$ always appear with an electric field $\partial_i\mu_R\new{-}E_{iR}$. This was discussed for the 1-CTP case in \cite{Akyuz:2023lsm} and extends to the 2-CTP case as well. 
\newline One way to see this would be to work entirely in terms of the $\phi$ fields first and then introduce background sources like above. The action would be KMS invariant by construction. We then use the fact that the right degrees of freedom are the three Nambu-Goldstone modes $(\phi_A,\phi_+,\phi_-)$ and the density $\mu_R$. In terms of these variables, $\partial_i\dot{\phi}_R=\partial_i\mu_R\new{-}E_{i,R}$. 
\newline On coupling to background sources, the fully gauge-invariant quadratic action is 
\begin{equation}
\begin{split}
\frac{4}{\chi}\mathcal L_{(2)} &= 
	4iT D (\nabla_i \phi^A)^2
	+ \mu^R \nabla_0 \phi^A - D\nabla_i\phi^A (\partial_i \mu^R  \new{-E_{i,R}}) \\
	& + \nabla_0 \phi^+ \nabla_0 \phi^- - D \partial_t \nabla_i\phi^+\nabla_i\phi^- 
	+ 2 D \partial_t\nabla_i  \phi^A \nabla_i\phi^-,
\end{split}
\end{equation}
From here, we can get the different densities by varying w.r.t the $A_0^I$ fields:
\begin{equation}
n^I = \chi \{ \mu^R, D \nabla^2 \phi^A, \nabla_0 \phi^+, \nabla_0 \phi^-\}    
\end{equation}
The non-zero propagators are
\begin{equation}\label{eq_2ctp_prop}
    \begin{split}
&G_{RR}=\frac{2T\chi q^2}{(\omega^2+q^4)}\\
&G_{RA}=\frac{\chi q^2}{4(\omega+iq^2)}\\
&G_{+-}=\frac{\chi q^2}{4(\omega+iq^2)}\\
& G_{+R}=\frac{T\chi q^2}{(\omega^2+q^2)}
    \end{split}
\end{equation}
where, we define $G_{IJ}=\frac{\delta\log{Z_4}}{(\delta iA_0^I)(\delta i A_0^J)}$ and $q_i=\sqrt{D}k_i$. 
Similarly, we also calculate different correlators. Our convention for the correlators is
\begin{equation}
    G_{IJKL}=\frac{\delta \log Z_4}{(\delta iA_{I0}(x_1))(\delta iA_{J0}(x_2))(\delta iA_{K0}(x_3))(\delta iA_{L0}(x_4))}
\end{equation}
where $t_1<t_2<t_3<t_4$.

\subsection{Correlators}\label{app_correlators}
As noted in section-\ref{ssec_scaling_obs}, not all $4^4$ correlators on the 2-CTP are independent, and there are only 6 independent real-valued correlators. We can pick a convenient basis of correlators in the 2-CTP EFT and express everything else in terms of these. We pick $(G_{RRRR},\, G_{--++},\, G_{-+-+},\,G_{RR-+},\, G_{R-+R},\, G_{-+RR})$ as our basis. Out of these, as we note in Appendix-\ref{app_scaling_corr}, $G_{--++}$ is the only correlator that cannot be obtained from 1-CTP correlators even at leading order. We write here the leading order expression for this correlator in momentum space. It turns out that in the particle-hole symmetric case, the leading order contribution comes only from the otoc parameters $\lambda_{1,2}$:
\begin{equation}\label{eq_app_Gmmpp}
\begin{split}
G_{--++}=&\frac{(4/\chi)^2}{D\beta}G_{-+}(p_1)G_{-+}(p_2)G_{+-}(p_3)G_{+-}(p_4)\\
\times\frac{1}{q_1q_2q_3q_4}&\Big[
  4\lambda_1\left(q_1q_2+q_3q_4\right)-\lambda_2\left((q_1+q_2)(q_3+q_4)\right)\Big]
\end{split}
\end{equation}
where $p_i\equiv (\omega_i,k_i)$, $q_i=\sqrt{D}k_i$ and $G_{+-}(p)$ is the propagator from \eqref{eq_2ctp_prop}. Unlike the 1-CTP case, the static limit of this correlator does not reduce to a lower-point function \cite{Delacretaz:2023ypv}, consistent with the fact that these parameters are beyond regular (1-CTP) transport parameters and cannot be captured through 1-CTP correlators:
\begin{equation}\label{eq_app_Kubo_freq}
\begin{split}
 \lim_{\omega_i\to 0}&G_{--++}(p_1,p_2,p_3,p_4)\\
 &\!\!=\frac{4\lambda_1(k_1 k_2+k_3 k_4)-\lambda_2(k_1+k_2)(k_3+k_4)}{\beta D^2k_1k_2k_3k_4}.
\end{split}
\end{equation}
We can take specific choices for the momenta to isolate these transport parameters. We note one such configuration in \eqref{eq_Kubo}.

On the other hand, as we discussed in the main text, certain OTOCs can be predicted entirely from TOCs - one example of this is the correlator $G_{R-+R}$. As we show in Appendix-\ref{app_scaling_corr}, at leading order, $G_{R-+R}$ is captured by a linear combination of $\text{Im}(g_2)$ (an entirely 2-CTP correlator) and $g_0$. Despite getting a leading order contribution  from $g_2$, its leading order expression gets fixed completely in terms of 1-CTP parameters $\sigma''$ and $\chi''$. In momentum space, it has the following expression,
\begin{equation}
    \begin{split}
        G_{R-+R}&=\frac{4}{D\chi^2}G_{RR}(p_1)G_{-+}(p_2)G_{+-}(p_3)G_{R+}(p_4)\\
        &\times\frac{q_1q_2}{q_1^2q_2^2q_3^2q_4^2}\Big(\sigma''\bigg[q_1 q_2 q_3\big(i q_4\big[(q_1^2+q_4^2)\\
        &\qquad-\left(q_2+q_3\right)\left(q_1+q_4\right)\big]\\
        &\qquad-2 q_4 \omega_2-\left(q_2-q_3+q_4\right)\omega_4\big)\\
        &\qquad-q_2 q_3 q_4\left(q_1+q_2-q_3\right)\omega_1\bigg]\\
        &\qquad+2 D\chi'' q_1 q_4\big(q_2^2 \omega_3-q_3^2 \omega_2+i \omega_2\omega_3\big)\Big)
    \end{split}
\end{equation}


\section{Scaling of the correlators}\label{app_scaling_corr}
In this section, we explain the scaling of correlators presented in Table-\ref{tab:scaling_correlators}. To get the scaling of $(g_0,\, g_1,\, g_2)$ in the hydrodynamic regime, we need to express them in terms of our EFT correlators. Since this is not a one-to-one map, we work backwards. We first express our EFT basis of correlators in terms $(g_0,\, g_1,\, g_2)$ using \eqref{eq_KeldyshRotation_Z4} and trace cyclicity. This map can then be inverted to express $(g_0,\, g_1,\, g_2)$ in terms of the EFT correlators. The most general expression is a complicated expression of the form
\begin{equation*}
    \begin{pmatrix}
       \text{Re}(g_0)\\
        \text{Re}(g_1)\\
         \text{Re}(g_2)\\
          \text{Im}(g_0)\\
           \text{Im}(g_1)\\
            \text{Im}(g_2)
\end{pmatrix}=F(e^{i\beta\partial_{t_1}},e^{i\beta\partial_{t_2}},e^{i\beta\partial_{t_3}})\begin{pmatrix}
        G_{RRRR}\\
        G_{--++}\\
        G_{-+-+}\\
        G_{RR-+}\\
        G_{R-+R}\\
        G_{-+RR}
    \end{pmatrix}
\end{equation*}
where $F$ is a $6\times 6$ matrix. But given that we want only the leading order expressions, we can expand in $\beta\partial_t$. On expanding, we get that at leading order in scaling, the regular time-ordered Wightman function in terms of the EFT correlators is, 
\begin{equation}
    \begin{split}
&\text{Re}(g_0)=\frac{1}{64}G_{RRRR}\left(1+\mathcal{O}(t^{-1})\right)\\
  &\text{Im}(g_0)=\frac{1}{32}\left(G_{-+RR}+G_{RR-+}\right)\\
  &\qquad-\frac{\beta}{1024}\left(9(\partial_{t_1}+\partial_{t_2})+2\partial_{t_3}\right)G_{RRRR}\left(1+\mathcal{O}(t^{-1})\right)
    \end{split}
\end{equation}
We can subtract the Wightman function from $g_1$ to capture subleading terms,
\begin{equation}
\begin{split}
   &\text{Re}(g_0-g_1)=\frac{\beta}{32}(\partial_{t_1}+\partial_{t_2})G_{RR-+}\left(1+\mathcal{O}(t^{-1})\right)\\
  &\text{Im}(g_0-g_1)=\frac{1}{16}G_{RR-+}\left(1+\mathcal{O}(t^{-1})\right)
  \end{split}
   \end{equation}
Similarly, we can get the true 2-CTP correlator $g_2$ 
\begin{equation}
    \begin{split}
&\text{Re}(g_2-g_0)=\frac{1}{16}G_{--++}\\
&-\frac{\beta}{256}\Big(\left(\partial_{t_1}+\partial{t_2}-2\partial{t_3}\right)G_{-+RR}\\
  &-\left(\partial{t_1}-3\partial{t_2}+2\partial{t_3}\right)G_{RR-+}\\
  &+4\left(3\partial{t_1}+\partial{t_2}+2\partial{t_3}\right)G_{R-+R}\Big)\\
&-\beta^2\Big(5\partial{t_1}^2+6\partial{t_1}\partial{t_2}+\partial{t_2}^2\\
&+14\partial{t_1}\partial{t_3}+22\partial{t_2}\partial{t_3}+8\partial{t_3}^2\Big)G_{RRRR}\left(1+\mathcal{O}(t^{-1})\right)
\end{split}
\end{equation}
And, 
\begin{equation}
    \begin{split}
       \text{Im}(g_0-g_2)=&\frac{1}{16}G_{R-+R}\\
       &+\frac{\beta}{512}(\partial_{t_1}+\partial_{t_2}+2\partial_{t_3})G_{RRRR}(1+\mathcal{O}(t^{-1}))     \end{split}
\end{equation}
These relations explain the scaling in Table-\ref{tab:scaling_correlators}. More precisely, from the EFT, we find that 
\begin{equation*}
    \begin{split}
  &G_{RRRR}\sim t^{-3d/2}\\
  &G_{RR-+}\sim t^{-(3d/2+1)}\\
  &G_{--++}\sim t^{-(3d/2+2)}
    \end{split}
\end{equation*}
The scaling for the $\beta=0$ case differs because $\lim_{\beta\to 0}G_{RR-+}=\mathcal{O}(t^{-3d/2+1})$ i.e., the leading order term vanishes (for all time orderings).


\section{Further numerical results}\label{app_num}

\begin{figure}
\includegraphics[width=1\columnwidth]{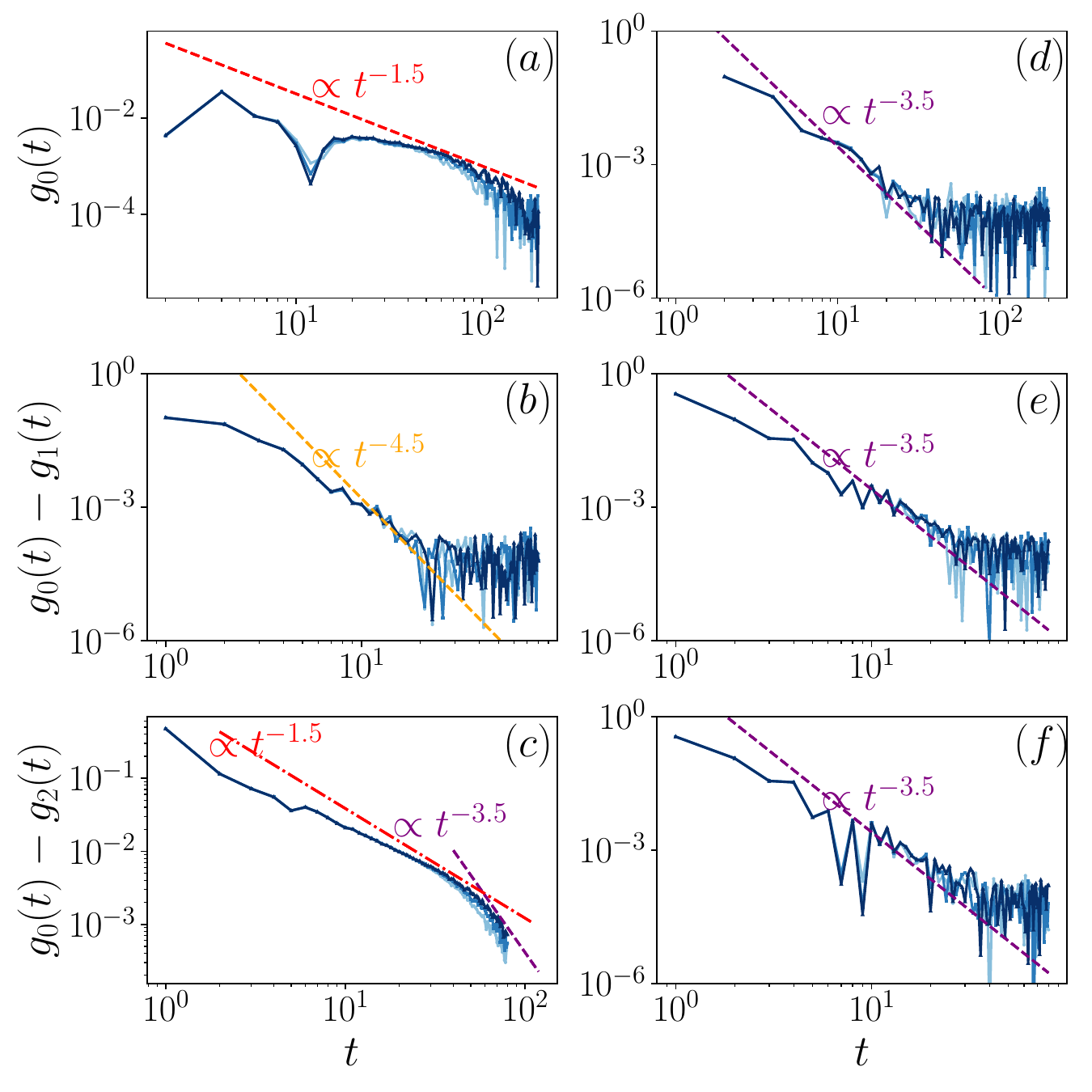}
\caption{Similar to Fig.~\ref{Fig-Ising-finiteT}, results in the mixed-field Ising model but at infinite temperature $\beta = 0.0$.
}
\label{Fig-Ising-ifiniteT}
\end{figure}

\begin{figure}
\includegraphics[width=1\columnwidth]{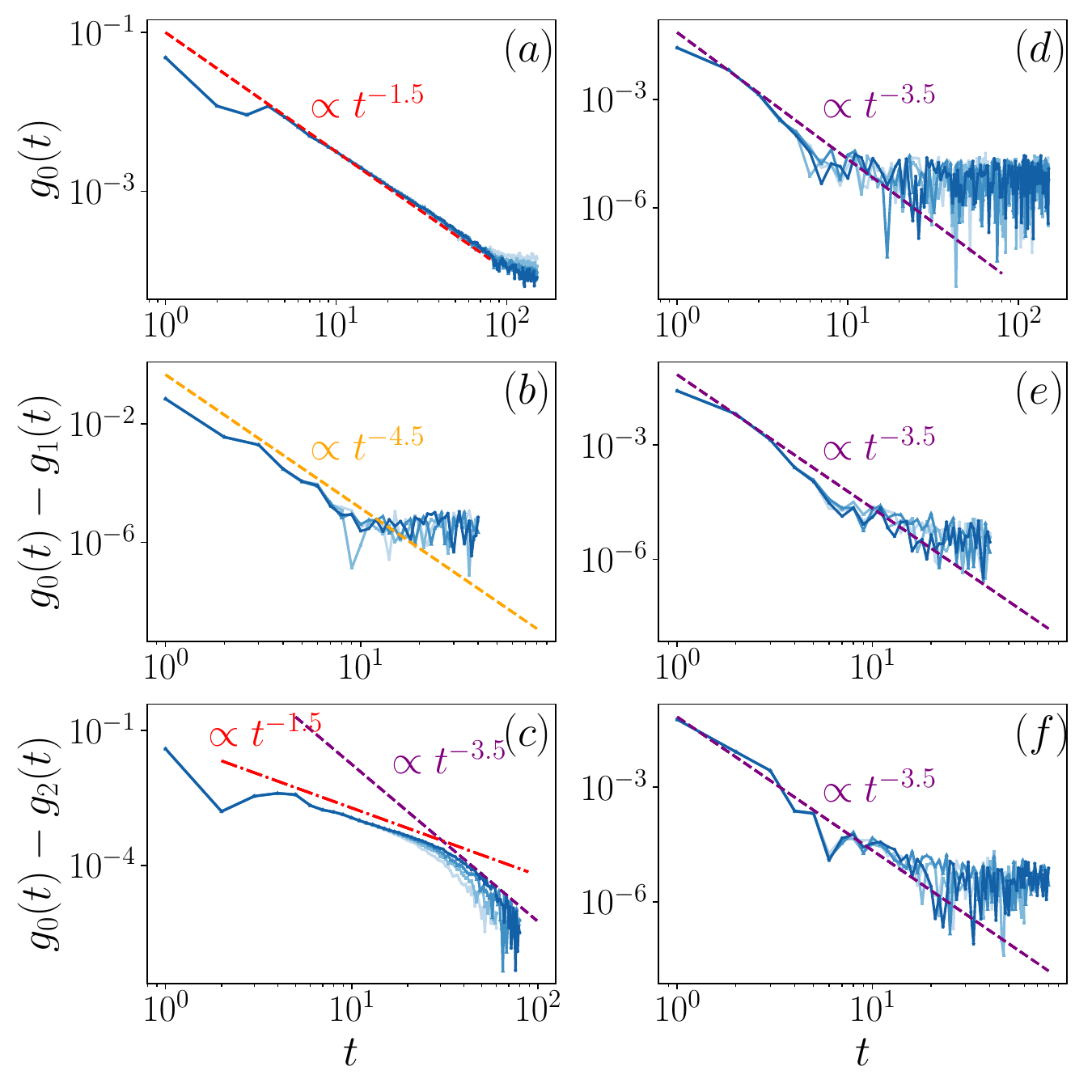}
\caption{Results in the Floquet XXZ model at $\mu = 0.0$:
$g_0(t)$ [(a)(d)]; $g_0(t)-g_1(t)$ [(b)(e)] and $g_0(t)-g_2(t)$ [(c)(f)].
The real and imaginary parts are presented in the left and right panels, respectively.
The dashed lines indicate EFT prediction in Table~\ref{tab:scaling_correlators}. The dash--dot line in panel (c) is shown as a guide to the eye. Here the system sizes are $L=24,26,28,30$ (from light to dark). 
The parameters are chosen as ${\cal J} = \frac{\pi}{8},\ {\cal J}^\prime = \frac{\pi}{4},\ \lambda = \new{-\frac{\pi}{8}}$. 
}
\label{Fig-FXXZ-infiniteT}
\end{figure}

\begin{figure}
\includegraphics[width=1\columnwidth]{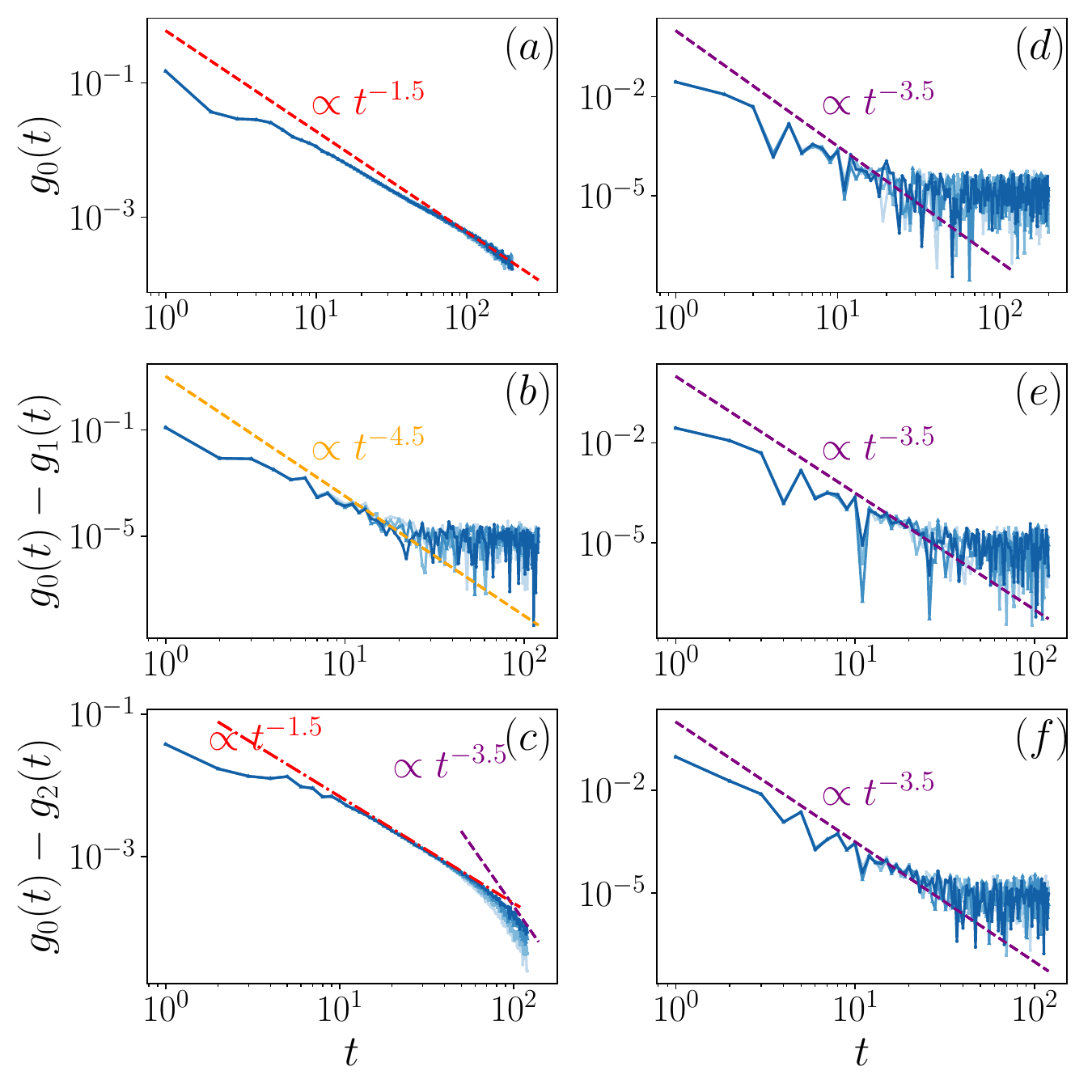}
\caption{Similar to Fig.~\ref{Fig-FXXZ-infiniteT}, results in the Floquet XXZ model, but for parameters ${\cal J} = \frac{\pi}{10},\ {\cal J}^\prime = \frac{\pi}{4},\ \lambda = \new{-\frac{\pi}{10}}$. 
}
\label{Fig-FXXZ-infiniteT10}
\end{figure}

In this appendix, we report the three independent 4-point functions discussed in Secs.~\ref{sec_EFT_properties} and \ref{sec_numerics} for various other choices of parameters and temperature, shown in Figs.~\ref{Fig-Ising-ifiniteT},~\ref{Fig-FXXZ-infiniteT} and ~\ref{Fig-FXXZ-infiniteT10}. 
In Fig.~\ref{Fig-Kappa3}, we also show the 3-point functions in the mixed field Ising model.

\begin{figure}[t]
\includegraphics[width=1\columnwidth]{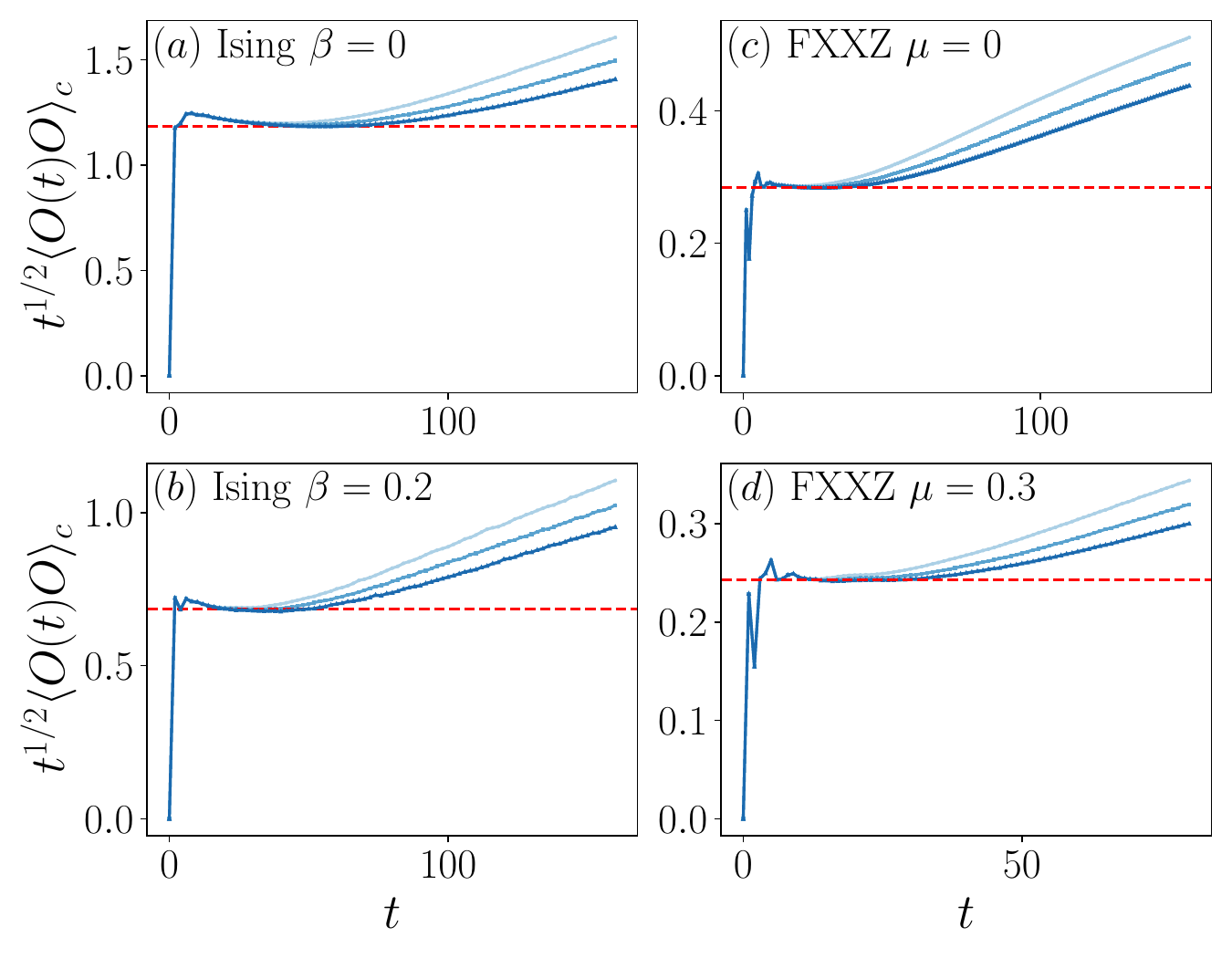}
\caption{Connected 2-point correlation function 
$t^{1/2}\langle O(t) O \rangle_{c}$ versus time $t$, 
where $O = o_n$ for the mixed-field Ising model and 
$O = \sigma^z_n$ for the Floquet XXZ model \new{for system sizes $L = 24,26,28$}. The dashed lines indicate the approximate plateau values as a guide to the eye. \new{They are related to the susceptibility as $\chi=A\sqrt{4\pi D}$ with $A$ the plateau values and $D$ the diffusion constant}. In [(c)(d)], the parameters are chosen as ${\cal J} = \frac{\pi}{8},\ {\cal J}^\prime = \frac{\pi}{4},\lambda = \new{-\frac{\pi}{8}}$. 
}
\label{Fig-LD}
\end{figure}

\begin{figure}
\includegraphics[width=1\columnwidth]{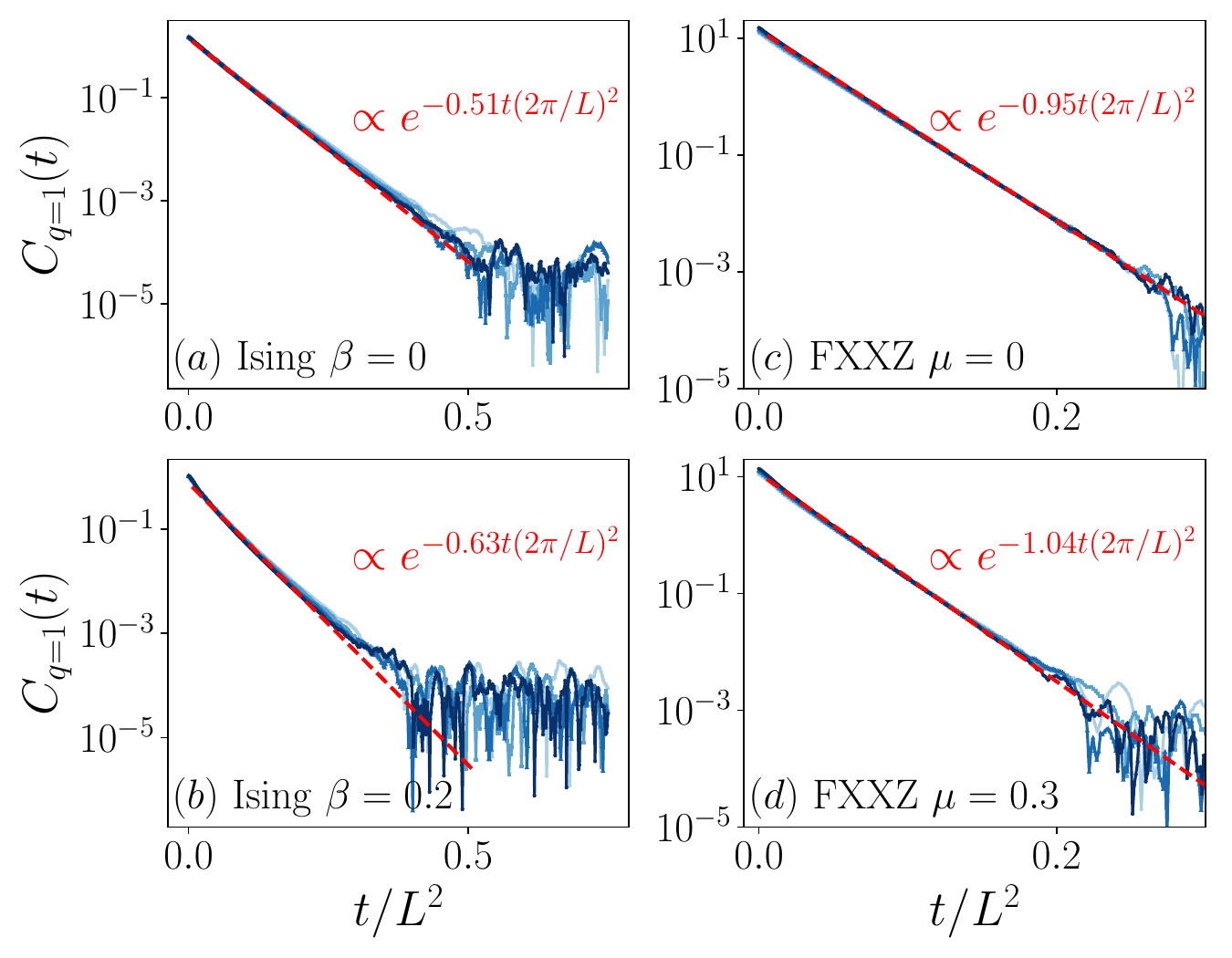}
\caption{
Auto-correlation function $C_{q}(t)=\langle O_{q}(t)O_{q}\rangle$ of density wave operator $O_{q}(t)=\sum_{n=0}^{L-1}\cos(\frac{2\pi nq}{L})o_{n}$. Here the longest wavelength is considered $q=1$. The dashed lines indicate the exponential fit. In [(c)(d)], the parameters are chosen as ${\cal J} = \frac{\pi}{8},\ {\cal J}^\prime = \frac{\pi}{4},\ \lambda = \new{-\frac{\pi}{8}}$. 
}
\label{Fig-DW}
\end{figure}

In addition to determining the charge susceptibility $\chi$, we also calculate the connected two-point correlation function of the conserved  local densities,
$o_n = h_n,\ \sigma^z_n$ for the mixed field Ising model and the Floquet XXZ model, respectively. 
Results are shown in Fig.~\ref{Fig-LD} where we show $t^{1/2}\langle o_{n}(t)o_{n}\rangle_{c}$ as a function of $t$. Furthermore, for the diffusion constant, we consider the density wave operator $  O_{q}(t)=\sum_{n=0}^{L-1}\cos(\frac{2\pi nq}{L})o_{n}
$. Here we study the longest wavelength $q=1$ and show the auto-correlation function $C_{q}(t)=\langle O_{q}(t)O_{q}\rangle$ in Fig.~\ref{Fig-DW}. \new{In summary, we find the following diffusivities and susceptibilities for the models studied:
\begin{align*}
&\hbox{Ising, $\beta=0$}& D&\approx 0.51 & \chi_T&\approx3.00\ (3.02) \\
&\hbox{Ising, $\beta=0.2$}& D&\approx 0.63& \chi_T&\approx1.93\ (2.09) \\
&\hbox{Floquet XXZ, $\mu=0$}& D&\approx 0.95 & \chi_T&\approx 0.98\ (1.00) \\
&\hbox{Floquet XXZ, $\mu=0.3$}& D&\approx 1.04 & \chi_T&\approx 0.88\ (0.92)
\end{align*}
where the susceptibilities $\chi_T\equiv \frac{\chi}{\beta}$ are defined using the variance,  
\begin{equation}
\begin{split}
   &\frac{\chi(\beta)}{\beta}=(\langle H^{2}\rangle_{\beta}-\langle H\rangle_{\beta}^{2})/L\\ &\frac{\chi(\mu)}{\beta}=(\langle S_{z}^{2}\rangle_{\mu}-\langle S_{z}\rangle_{\mu}^{2})/L\ .
   \end{split}
\end{equation}
and the $\beta=0$ case shows the value that $\frac{\chi}{\beta}$ saturates to since $\lim_{\beta\to 0}\chi=0$. The parentheses show susceptibilities obtained directly from the calculating the variance numerically while the other values are obtained from the plateau values in the connected two-point functions in Fig. \ref{Fig-LD}. Note that the (infinite temperature) spin susceptibility of the Floquet XXZ model can be easily obtained analytically: $ Z(\mu) \equiv  \Tr e^{\mu S_z} \propto \cosh \mu$ so that $\chi(\mu) = d^2 \log Z / d \mu^2 = 1/\cosh^2(\mu)$. Similarly, for the mixed field Ising model at infinite temperature, $\frac{\chi}{\beta}=J^2+g_x^2+g_z^2$.}


\section{Comment on infinite temperature correlators}

We observe an apparent discrepancy between the EFT predictions and some of the numerics for $\beta=0$, most visible in the right panels in Fig.~\ref{Fig-Ising-ifiniteT} and \ref{Fig-FXXZ-infiniteT10}: the data seems to show slower decay consistent with the $\beta>0$ EFT predictions rather than the $\beta=0$ ones. These power laws are all very fast, making them notably difficult to observe in simulations of quantum many-body systems \cite{Matthies:2024lqx}; furthermore, the disagreement is less or not apparent for different choices of parameters, see Fig.~\ref{Fig-FXXZ-infiniteT}. It is thus possible (and perhaps likely) that larger scale simulations would resolve this discrepancy. Nevertheless, for completeness and to help guide future studies, we document this possible tension in this section. 

\begin{figure}
\includegraphics[width=1\columnwidth]{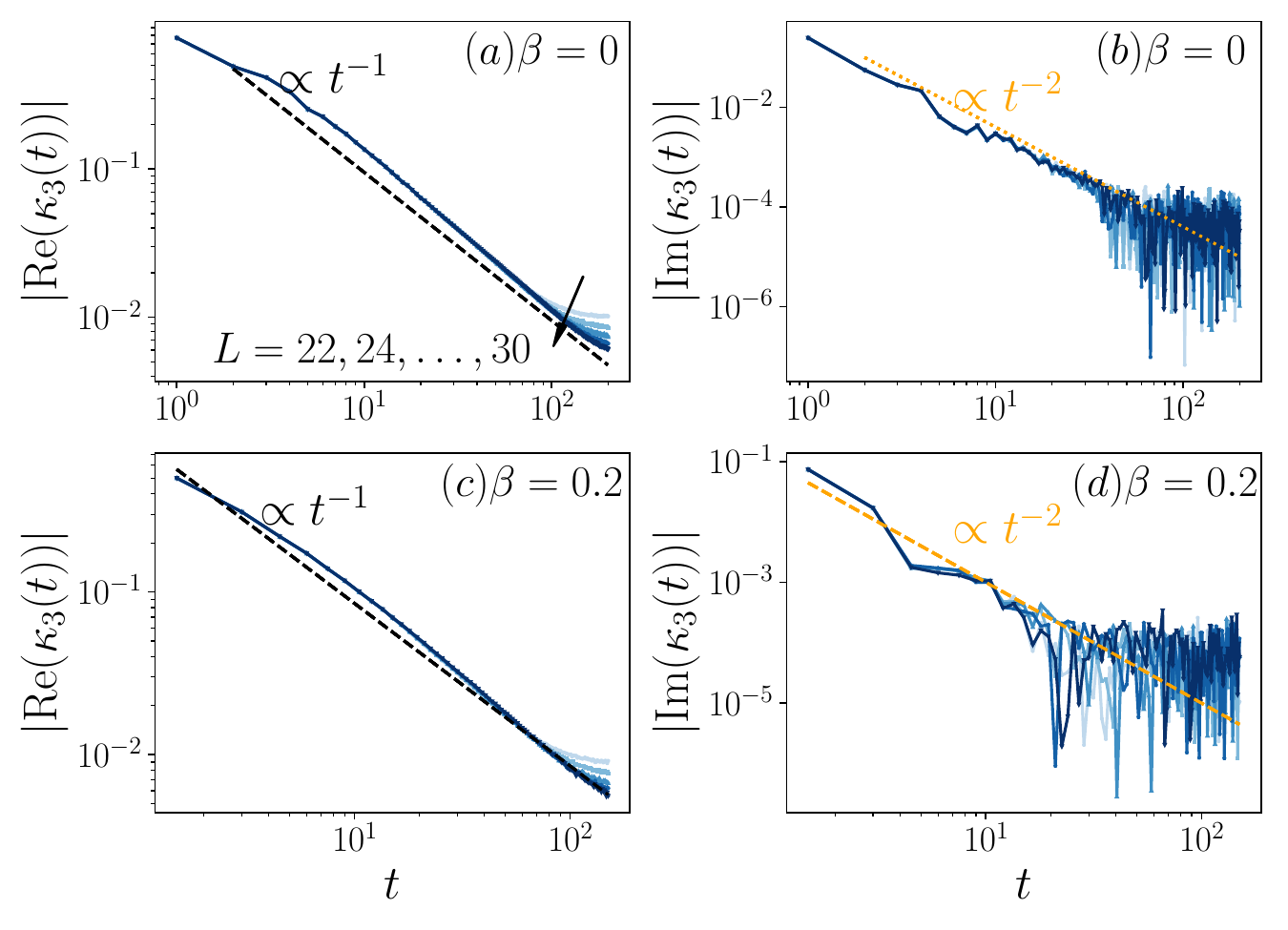}
\caption{Connected three-point correlation functions in the mixed field Ising model at infinite temperature $\beta = 0.0$ and [(a)(b)] and finite temperature $\beta = 0.2$ [(c)(d)]. The dashed and dotted line indicate the scaling $\propto t^{-1}$ and $\propto t^{-2}$, respectively.
}
\label{Fig-Kappa3}
\end{figure}

This discrepancy is not tied to OTOCs, and is visible in correlators representable on a 1-CTP as well. The simplest observable that manifests it is the 3pt function. Trace cyclicity implies that there are only two independent real 3pt functions of identical operators: the real and imaginary parts of 
\begin{equation}
\kappa_3(t) = \Tr \left(\rho\,\mathcal O(2t) \mathcal O(t) \mathcal O(0)\right) \, .
\end{equation}
These are shown in Fig.~\ref{Fig-Kappa3} for the Hamiltonian spin chain. As explained below, the 1-CTP EFT predicts that $\kappa_3$ is mostly real and decays as $1/t$ at both $\beta=0$ and $\beta>0$, in agreement with the numerics in Fig.~\ref{Fig-Kappa3}. Let us now turn to the subleading imaginary part: the EFT predicts $\Im \kappa_3\sim 1/t^2$ at $\beta>0$, which seems to agree with Fig.~\ref{Fig-Kappa3}(d), but predicts a faster decay $\Im \kappa_3\sim 1/t^{>2}$ at $\beta=0$, which is not observed in Fig.~\ref{Fig-Kappa3}(b). This apparent mismatch between fluctuating hydrodynamic predictions and $\beta=0$ numerics therefore arises in conventional higher-point functions as well.

In the rest of this section, we explain how the EFT prediction for the scaling is obtained. In terms of the fields in Keldysh basis, the two correlators are
\begin{equation}
    \begin{split}
       & \text{Re}(\kappa_3(t))= G_{rrr}(1+O(t^{-1}))\\
       &\text{Im}(\kappa_3(t))=G_{rar}(1+O(t^{-1}))
    \end{split}
\end{equation}
These correlators were studied in \cite{Delacretaz:2023ypv}, and it was shown that at leading order, they are completely fixed in terms of $\chi'$ and $D'$ with the scaling
\begin{equation}\label{eq_Grrr_Grar}
\begin{split}
  &G_{rrr}\sim \frac{\chi T}{(Dt)^{d}}\left(c+O(t^{-1})\right)\\
  &G_{rar}\sim \frac{\chi }{(Dt)^{d}} \frac{1}{t}\left(c'+O(t^{-1})\right)
\end{split}
\end{equation}
where the dimensionless coefficients $c,\,c'$ are linear combinations of the dimensionless ratios $\frac{T (d\chi/d\mu)}{\chi}$ and $\frac{T (dD/d\mu)}{D}$. Note that $T\chi \equiv  T dn/d\mu$ and $D$ typically have finite limits in lattice models with bounded local Hilbert space as $\beta\to 0$ (as do $c$ and $c'$). Thus, the two correlators are related by the scaling
\begin{equation}
\begin{split}
&G_{rar}\sim (\beta\omega +(\tau\omega)^m)G_{rrr}\\
&\lim_{\beta\to 0}G_{rar}\sim (\tau\omega)^m G_{rrr}
\end{split}
\end{equation}
where $\tau$ is some timescale that comes from higher-order terms in the action. Eq.~\eqref{eq_Grrr_Grar} implies that $m\geq 2$. This explains why the EFT predicts a subleading scaling for the $\beta=0$ case. 
Determining the timescale $\tau$ and the precise value of $m$ would require a more systematic analysis of higher-order terms in the EFT than presented here. Here we illustrate one possible mechanism by which such a timescale may emerge.

In the 1-CTP theory, there are three kinds of cubic vertices possible: $c_1\phi_a n^2$, $c_2 \phi_a^2 n$ and $c_3 \phi_a^3$ each with additional derivative structure. At a fixed order in derivatives, the coefficients have different dimensions related by $[c_1]=[\chi\omega]^{-1}[c_2]=[\chi^2\omega]^{-2}[c_3]$ since $[n]=[\chi][\phi_a]$. These vertices contribute as follows in momentum space, 
\begin{equation}
\begin{split}
 &  G_{rrr}\sim c_1\frac{(\chi T)^2}{\omega^3}+c_2\frac{(\chi T)}{\omega^3}+c_3\frac{1}{\omega^3}\\
 & G_{rar}\sim c_2\frac{\chi}{\omega^2}
\end{split}
\end{equation}
There are two possible ways in which $\lim_{\beta\to 0}G_{rrr}$ can remain finite. The first is if the coefficients scale as $c_i\sim\mathcal{O}(\beta^0)$ in which case $\lim_{\beta\to 0}G_{rar}=0$, no timescale constructed from these coefficients survives in the $\beta\to 0$ limit. The second possibility is that the coefficients scale as $c_i\sim \mathcal{O}(\beta^{-1})$ with their values tuned such that the singular contributions to $G_{rrr}$ cancel while $\lim_{\beta\to 0}G_{rar}$ is finite. In this case, a time scale $\tau$ emerges that is finite at $\beta=0$. At higher orders in the derivative expansion, there are more possibilities for coefficients that lie beyond classical transport coefficients \cite{Jain:2020zhu} and can make such a cancellation possible. One may expect that the relations among the coefficients $c_i$ required for such cancellation arise as a consequence of KMS.

\bibliography{otoceft}{}

\end{document}